\documentclass{article}
\usepackage{PRIMEarxiv}

\usepackage[utf8]{inputenc} 
\usepackage[T1]{fontenc}    
\usepackage{url}            
\usepackage{amsfonts}       
\usepackage{nicefrac}       
\usepackage{microtype}      
\usepackage{fancyhdr}       

\usepackage[square,numbers]{natbib}

\usepackage{amsmath,amsfonts,bm}

\def\Figref#1{Figure~\ref{#1}}

\def\eqref#1{equation~\ref{#1}}

\def\1{\bm{1}}

\DeclareMathAlphabet{\mathsfit}{\encodingdefault}{\sfdefault}{m}{sl}
\SetMathAlphabet{\mathsfit}{bold}{\encodingdefault}{\sfdefault}{bx}{n}

\def\gG{{\mathcal{G}}}
\def\gH{{\mathcal{H}}}

\def\sA{{\mathbb{A}}}

\def\sF{{\mathbb{F}}}

\def\sV{{\mathbb{V}}}

\usepackage{graphicx}
\usepackage{xcolor}
\definecolor{lightblue}{RGB}{31,119,180}
\definecolor{navy}{RGB}{0,0,128}
\usepackage[colorlinks=true,
            linkcolor=navy,
            citecolor=navy,
            urlcolor=blue]{hyperref}

\usepackage{enumitem} 

\usepackage{booktabs}

\usepackage{caption}
\usepackage{subcaption}

\usepackage{tikz}
\usetikzlibrary{shapes}
\usetikzlibrary{plotmarks}

\definecolor{repr}{RGB}{255,127,14}
\definecolor{sub}{RGB}{44,160,44}
\definecolor{mult}{RGB}{214,39,40}
\definecolor{molhgcn}{RGB}{31,119,180}

\usepackage{xcolor,pifont}
\newcommand*\colourcheck[1]{%
  \expandafter\newcommand\csname #1check\endcsname{\textcolor{#1}{\ding{52}}}%
}

\usepackage{enumitem} 

\colourcheck{sub}
\title{A molecular hyper-message passing network with functional group information}

\author{
  Fangying Chen*, Junyoung Park*, Jinkyoo Park \\
  KAIST \\
  Daejeon, Korea \\
  \texttt{\{gyjin32, junyoungpark, jinkyoo.park\}@kaist.ac.kr} \\
}

\begin{document}
\maketitle

\begin{abstract}
We proposed the molecular hyper-message passing network ({\fontfamily{lmtt}\selectfont MolHMPN}) that predicts the properties of a molecule with prior knowledge-guided subgraph. Modeling higher-order connectivities in molecules is necessary as changes in both the pair-wise and higher-order interactions among atoms results in the change of molecular properties. Many approaches have attempted to model the higher-order connectivities. However, those methods relied heavily on data-driven approaches, and it is difficult to determine if the utilized subgraphs contain any properties of interest or have any significance on the molecular properties. Hence, we propose {\fontfamily{lmtt}\selectfont MolHMPN} to utilize the functional group prior knowledge and model the pair-wise and higher-order connectivities among the atoms in a molecule.
Molecules can contain many types of functional groups, which affect the properties the molecules.  For example, the toxicity of a molecule is associated with toxicophores, such as nitroaromatic groups and thiourea. {\fontfamily{lmtt}\selectfont MolHMPN} uses functional groups to construct hypergraphs, modifies the hypergraph using domain knowledge-guided modification scheme, embeds the graph and hypergraph inputs using a hypergraph message passing ({\fontfamily{lmtt}\selectfont HyperMP}) layer, and uses the updated graph and hypergraph embeddings to predict the properties of the molecules. Our model provides a way to utilize prior knowledge in chemistry for molecular properties prediction tasks, and balance between the usage of prior knowledge and data-driven modification adaptively. We show that our model is able to outperform the other baseline methods for most of the dataset, and show that using domain knowledge-guided data-learning is effective. 
\end{abstract}

\section{Introduction}
Toxicological screening is vital for the development of new drugs, the evaluation of the therapeutic potential of existing molecules, and the assessment of pharmacological activity and toxicity potential of new molecules on human. Traditionally, toxicity studies of molecules relied on animal testing, which can provide inadequate bases for predicting clinical outcomes on humans \cite{anitest:2015}. The U.S. Food and Drug Administration (FDA) has also estimated that it takes more than eight years to test and study a new drug before its approval to the general public, which includes early laboratory and animal testing \cite{FDA:2015}. Machine learning (ML) methods have therefore been utilized widely to assess the effects that chemicals have on humans and the environment as it is able to utilize data with large data sizes, while reducing the time and cost it takes for drugs approval, and avoiding costly late-stage failures. 

In ML, graph neural networks (GNNs) have been used actively in molecule-related tasks for their ability to represent molecules as graphs. Representing molecules as graphs is natural and preferred since the molecular structure is inextricably linked to the molecular properties. In the graphs, the atoms and bonds of the molecules are represented by the nodes and edges of the graphs. These methods take the graphs as inputs and use the node features to predict the molecular properties. The connectivities between the nodes in the graphs include the pair-wise connectivities between two nodes that are connected by an edge, and the higher-order connectivities between nodes that are further apart.

To model the pair-wise connectivities in the graphs, the message passing neural network (MPNN) \cite{mpnn:2017}, which is a representative GNN architecture, has been devised as a fast simulation method to replace computationally expensive quantum mechanical simulations. Its variants \cite{dmpnn:2019, cmpnn:2020} have also shown their potentials in molecular properties prediction tasks. 
These pair-wise methods can model higher-order connectivities by stacking multiple MPNN layers. This, however, can cause the model the to suffer from oversmoothing \cite{dropedge:2020} or oversquashing \cite{alon2020bottleneck} problems.
Alternatively, the higher-order connectivities can be modeled by augmenting substructures, such as introducing virtual nodes \cite{agcn:2018} or combining multiple nodes \cite{gaan:2019, wgj:2020, gmeta:2020}. Similarly, hypergraphs contains hyperedges that are made up of nodes from a subgraph \citep{hgnn:2018, hcha:2021}. 
Examples of such subgraphs include frequently-occurring substructures \cite{wgj:2020}, K-hop neighbor substructures \cite{gmeta:2020}, and residual substructures that are unspecified by the graph adjacency matrix \cite{agcn:2018}. 
However, these methods have focused exclusively on data-driven approaches and it is hard to determine if those subgraphs contain any properties of interest or have any significance on the molecular properties. 

\begin{figure}[h]
\begin{center}
\includegraphics[width=9cm]{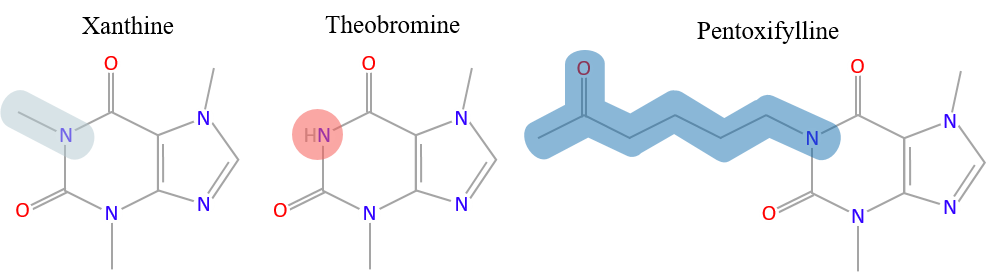}
\end{center}
\caption{Molecules of similar structures but different properties. Xanthine is found in caffeine and temporarily prevents or reduces drowsiness, theobromine is found in cacao and has mood improving effect, and pentoxifylline is a drug used to treat muscle pain in people with peripheral artery disease. The colored parts shows their difference. The grey and red parts show that the pair-wise interactions between two atoms can change the properties of the molecules, and the grey/red and blue parts show that the higher-order interactions between atoms can change molecular properties.}
\label{fig:molcompare}
\end{figure}

In chemistry, molecules are constructed from a carbon skeleton, onto which functional groups are attached to. The carbon skeleton is a chain of carbon atoms and is relatively unreactive. On the other hand, functional group is a group of atoms that are bonded together in a particular fashion, and determines the properties of the molecules (e.g. solubility, reactivity and lipophilicity) \cite{blackman2019chemistry}. 
Figure \ref{fig:molcompare} shows examples of molecules that have similar structures but with different properties. From Figure \ref{fig:molcompare}, the difference in the number of methyl, amine and ketone groups has resulted in the different effects that the compounds have on the human body \cite{kotera2008functional}; the changes in the pair-wise and higher-order interactions among the atoms can change the properties of the molecules. Hence, accounting for both pair-wise and higher-order interactions among atoms is essential for molecular properties prediction. The current study aims to incorporate the prior knowledge of functional groups to model the higher-order interactions in the molecule to ensure that the subgraphs are significant to the molecular properties of the molecules. The incorporation of prior knowledge to neural networks (NNs) has been attempted by several methods for their specific tasks \cite{raissi2019physics, park2019physics, long2018pde, yang2021hit}. 
However, it is difficult to determine which prior knowledge to exploit in closed form for the learning pipelines, and the wrong selection of prior knowledge can deteriorate the performances of the learning models. Therefore, ML models that can leverage the prior knowledge \textit{partially} and overcome the potentially unsuitable prior knowledge are needed.

In this paper, we propose a molecular hyper-message passing network ({\fontfamily{lmtt}\selectfont MolHMPN}) that is able to predict the properties of a molecule with prior knowledge-guided substructures. Our model ({\fontfamily{lmtt}\selectfont MolHMPN}) predicts the molecular properties by conducting the following sequential operations:

\begin{itemize}[leftmargin=*]
    \item \textbf{Constructing hypergraphs using functional groups.} Given a graph representation of a molecule that is constructed from its simplified molecular-input line-entry system (SMILES) string, {\fontfamily{lmtt}\selectfont MolHMPN} constructs the hyperedges according to functional groups that have been identified by chemists to represent the higher-order connectivities among the atoms. 
    Each hyperedge represents a functional group that is present in a molecule. The hyperedges can also be extended up to their $K$-hop neighborhood. 
    \item \textbf{Embedding the graph and hypergraph using hypergraph message passing layer ({\fontfamily{lmtt}\selectfont HyperMP}).} The {\fontfamily{lmtt}\selectfont HyperMP} consists of an atom graph convolution (AtomGC) and a functional group graph convolution (FuncGC) for the graphs and hypergraphs respectively. It performs message passing on the graphs and hypergraphs sequentially.
    \item \textbf{Modifying the hypergraph using the computed embeddings.}
    {\fontfamily{lmtt}\selectfont MolHMPN} modifies the input hypergraph by considering the original graph and hypergraph, and their respective embedded representations. This process updates the prior knowledge (i.e., input hypergraphs) with observations (i.e., embeddings) similar to that of the Bayesian approaches.
    \item \textbf{Predicting the molecular properties from the modified hypergraph.} {\fontfamily{lmtt}\selectfont MolHMPN} applies {\fontfamily{lmtt}\selectfont HyperMP} again to compute the embedding with the original graph and modified hypergraph, and predict the target label with the updated embeddings.
\end{itemize}

The key contribution of the current study is on the adaptation of functional groups using prior knowledge and the utilization of the prior knowledge selectively when conducting the molecular prediction tasks. 
Our novelties are summarized as follows:
\begin{itemize}[leftmargin=*]
\item \textbf{Providing a way to utilize the prior knowledge.}  {\fontfamily{lmtt}\selectfont MolHMPN} translates functional groups, which are based upon prior knowledge in chemistry, into hyperedges to process higher-order connectivities in molecules effectively. 
\item \textbf{Balancing between prior knowledge and data-driven scheme.} Without heavily relying on the functional group prior knowledge, {\fontfamily{lmtt}\selectfont MolHMPN} modifies such information adaptively depending on the target input. This can alleviate risk of using unsuitable information or representations of the target molecule.
\end{itemize}

We evaluate the effectiveness of {\fontfamily{lmtt}\selectfont MolHMPN} on several datasets that are used for molecular properties classification and regression tasks, and show that {\fontfamily{lmtt}\selectfont MolHMPN} is able to outperform the other baseline methods for most datasets. We also analyze the usage of different types of substructures and the effectiveness of the prior knowledge-guided data-driven modification for the prediction tasks.

\section{Methodology}
\begin{figure*}[h]
\begin{center}
\includegraphics[width={1.0\linewidth}]{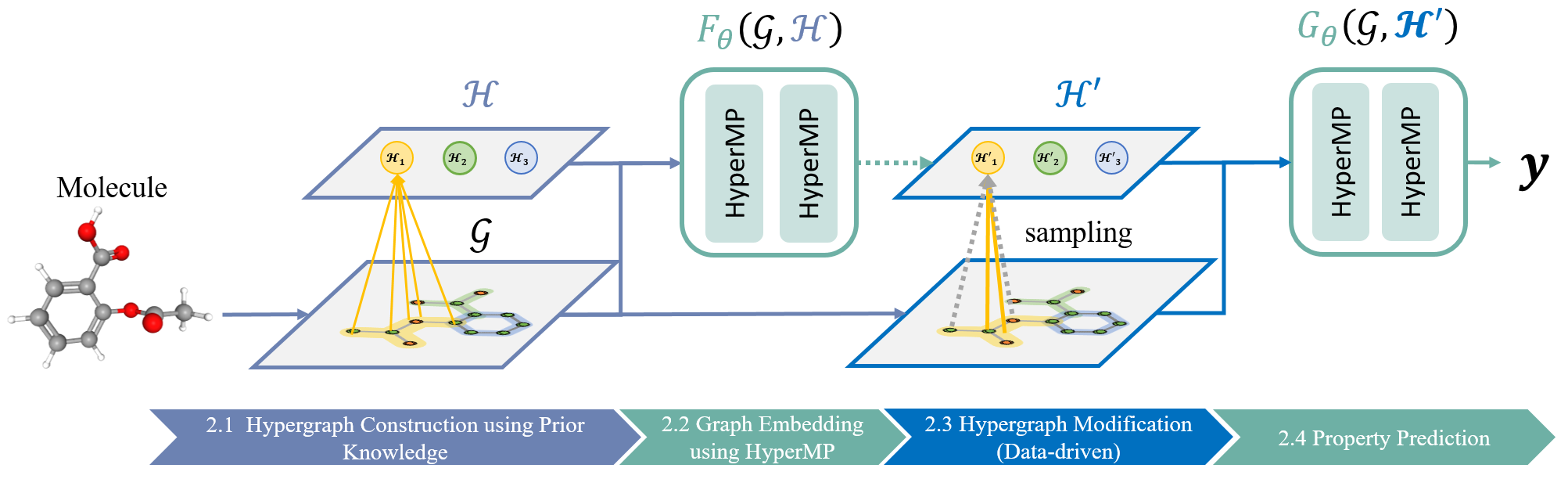}
\end{center}
\caption{\textbf{Overall architecture of {\fontfamily{lmtt}\selectfont MolHMPN}} }
\label{fig:molhmpn}
\vspace{-0.2cm}
\end{figure*}

This section highlights the methodology of the proposed {\fontfamily{lmtt}\selectfont MolHMPN}. In {\fontfamily{lmtt}\selectfont MolHMPN}, the hypergraphs are first constructed using the prior knowledge of functional groups and extended up to their $K$-hop neighborhood. The graph and constructed hypergraphs are then embedded using the {\fontfamily{lmtt}\selectfont HyperMP} layer(s) so as to modify the membership of the hyperedges using the computed embeddings. The graph and modified hypergraphs are then embedded again using the {\fontfamily{lmtt}\selectfont HyperMP} layer(s) to predict the target label with the updated embedding. \Figref{fig:molhmpn} shows the overall architecture of {\fontfamily{lmtt}\selectfont MolHMPN}.

\subsection{Hypergraph construction} 
\label{subsection:hg_construction}
Inspired by the significance of functional groups on the molecular properties as discussed in section 1, we utilize the knowledge of functional groups that are defined by chemists to let the model identify the similarities and differences of the molecules more easily. We represent the molecules as conventional pair-wise graphs and hypergraphs. The  conventional pair-wise graphs are defined as $\gG=\{\sV,\mathbb{E}\}$, where $\sV = \{v_1, ..., v_n\}$ is a set of $n$ nodes (atoms), and $\mathbb{E} \subset \sV \times \sV$ is a set of edges (bonds). The edge $e_{ij}$ exists if a bond between $v_i$ and $v_j$ exists. The features of $v_i$ and $e_{ij}$ are defined as $x_i$ and $x_{ij}$ respectively. The hypergraph is defined as $\gH = \{\mathcal{H}_k | k=1, ... , n_K\}$, where $\mathcal{H}_k$ is $k^\text{th}$ hyperedge that has a set of nodes as its members. The features of $\mathcal{H}_k$ are defined as $z_k$.

When constructing $\mathcal{H}$, atoms in cyclic and acyclic (open-chain) groups are considered separately. The minimal collection of cycles in the molecules are extracted as $\gH_k$. For the acyclic groups, the vicinity of the functional group is considered when extracting the hyperedge representation, which is defined as the central atom and the atoms that are attached to it \citep{kotera2008functional}. The main atoms that are used are carbon (C), nitrogen (N), oxygen (O), phosphorus (P) and sulfur (S), and the main bond types that are used are the single ($-$), double ($=$) and triple bonds ($\equiv$). The extraction process of the acyclic groups can be described as follows:

\begin{enumerate}
    \item Find a central atom (e.g., C, N, O, P or S) from $\gG$ and set it as $v_c$.
    \item Find the 1-hop neighborhood set $\sF_1(v_c)$ of $v_c$, which is given as $\sF_1(v_c)=\{v_j \in \mathcal{N}(v_c) \,|\, t(v_j) \in \sA_t, t(e_{ij}) \in \mathbb{B}_t\}$, where $\mathcal{N}(v_c)$ is the neighborhood of $v_c$, $t(\cdot)$ denotes the types of atom/bond, and $\sA_t, \mathbb{B}_t$ are the sets of target atom and bond respectively that are based accordingly to the target functional group. 
    \item Find the 2-hop neighborhood set $\sF_2(v_c)$ of $v_c$, which is given as $\sF_2(v_c)=\{v_k \in \bigcup\limits_{v_j \in \mathcal{N}(v_i)} \mathcal{N}(v_j) \, |\, t(v_j) \neq \text{C} \lor t(e_{ij}) \neq -\}$.
    \item The extracted hyperedge is hence $\mathcal{H}_k = \{v_c\} \cup \sF_1(v_c) \cup \sF_2(v_c)$.
\end{enumerate}
Different combinations of the central atoms, and $\sA_t$, $\mathbb{B}_t$ are used to match each functional group. Here, the prior knowledge of functional groups is applied in $\sA_t$ and $\mathbb{B}_t$. The remaining atoms that do not belong to any of the specified functional groups are put into the same hyperedge if they are connected by an edge. The 1-hop and 2-hop neighborhood sets are defined as changes in atom types within this range gives different functional groups. \Figref{fig:fgextract} shows an example of the hyperedge construction for the carboxyl group in aspirin. The list of functional groups used in this paper are given in Appendix \ref{appen:hyperconstr}. 

As we want to let the model leverage the prior knowledge (i.e. functional group) \textit{partially} and overcome potentially unsuitable prior knowledge, we consider an extension of the hyperedges so that a much higher-order interaction can be captured by the hyperedges. In that regard, the hyperedges are extended up to their $K$-local neighborhood as follows:
\begin{flalign}
    \label{eq:fg-extend}
    \gH_k=\bigcup\limits_{v_i \in \gH_k}\mathcal{N}_K(v_i)
\end{flalign}
where $\mathcal{N}_K(v_i)$ is the $K$-hop neighborhood set of $v_i$. This extension allows the hyperedge modification in Section \ref{subsection:learning_subgraph} to guided by the prior knowledge while being restricted to the scope of the hyperedge in the extended $\gH_k$. $\gG$ and $\gH$ will then be fed into the {\fontfamily{lmtt}\selectfont HyperMP} layer to compute the embedding that are needed to adjust the members of $\gH_k$.

\begin{figure}[t]
\centering
\begin{subfigure}{.18\textwidth}
  \centering
  \includegraphics[width=0.8\linewidth]{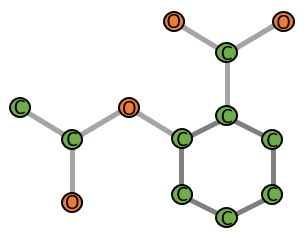}
  \caption{$\gG$}
  \label{fig:sub1}
\end{subfigure}%
\begin{subfigure}{.18\textwidth}
  \centering
  \includegraphics[width=0.8\linewidth]{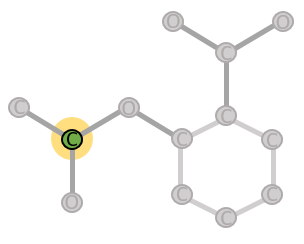}
  \caption{Set $v_c$}
  \label{fig:sub2}
\end{subfigure}
\begin{subfigure}{.18\textwidth}
  \centering
  \includegraphics[width=0.8\linewidth]{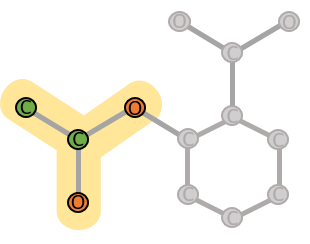}
  \caption{Find $\sF_1(v_c)$ }
  \label{fig:sub2}
\end{subfigure}
\begin{subfigure}{.18\textwidth}
  \centering
  \includegraphics[width=0.8\linewidth]{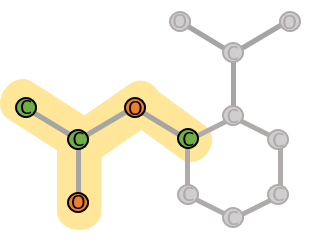}
  \caption{Find $\sF_2(v_c)$}
  \label{fig:sub2}
\end{subfigure}
\begin{subfigure}{.18\textwidth}
  \centering
  \includegraphics[width=0.8\linewidth]{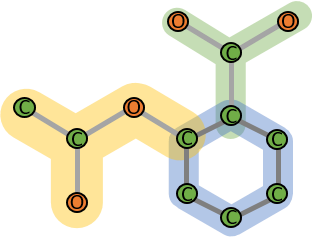}
  \caption{Extracted $\mathcal{H}$}
  \label{fig:sub2}
\end{subfigure}
\caption{\textbf{Hypergraph construction for aspirin.} a) $\gG$ of aspirin. b) Set carbon as $v_c$. c) To find $\sF_1(v_c)$, set $v_c -$ O, $v_c =$ O and $v_c -$ C, where $\{\text{O}, \text{C} \in \sA_t\}$ and  $\{-, = \in \mathbb{B}_t\}$. d) To find $\sF_2(v_c)$, find $v_j$ that is not $C$ and $e_{ij}$ that is not a single bond. e) All the extracted $\mathcal{H}_k$ of $\gG$.}
\label{fig:fgextract}
\end{figure}

\subsection{Graph and hypergrph embedding with Hypergraph message passing}
\label{subsection:hyper_mp_layer}
Modeling both the pair-wise (atom/bond) and higher-order (functional group) connectivities is crucial for conducting the molecule property predictions. Hence, we introduce the hypergraph message passing {\fontfamily{lmtt}\selectfont HyperMP} layer to integrate the information from both the atoms and functional groups. {\fontfamily{lmtt}\selectfont HyperMP} updates the input graphs via two steps: atom graph convolution (AtomGC) and functional group graph convolution (FuncGC). The general equation of the {\fontfamily{lmtt}\selectfont HyperMP} can be defined as:
\begin{align}
    \label{eq:hypermp}
    \gG', \gH' = \text{HyperMP}(\gG, \gH)
\end{align}
where $\gG'$ and $\gH'$ are the updated graph and hypergraph respectively.

\paragraph{AtomGC.} AtomGC is designed to model the pair-wise interactions between atoms that are bonded together. It involves updating the edge features using the features of the edges and nodes that it connects, and updating the node features using the updated edge features. The edge update step is given as:
\begin{align}
    \label{eq:atom-level-edge-update}
    x'_{ij} = f_{\text{bond}}(x_i, x_j, x_{ij})
\end{align}
where $x'_{ij}$ is the updated edge feature and $f_{\text{bond}}(\cdot)$ is the edge multi-layer perceptron (MLP). It is noteworthy that, for the target tasks, the edge information is essential as the chemical bonds contains crucial information about the molecular properties. In the node update step, the updated node features $x'_i$ is computed with $x'_{ij}$ as follows: 
\begin{align}
    \label{eq:atom-level-node-update}
    \alpha_{ij} &= f_{\text{attn}}(x_i, x_j, x_{ij}) \\
    x'_{i} &= f_{\text{atom}}\Big(x_i, \sum_{j=\mathcal{N}(i)}\alpha_{ij} x'_{ij}\Big)    
\end{align}
where $\alpha_{ij}$ is the attention coefficient of $e_{ij}$, $f_{\text{attn}}(\cdot)$ is the attention multi-layer perceptron (MLP) whose output activation is the sigmoid activation function, and $f_{\text{atom}}(\cdot)$ is the node MLP and $\mathcal{N}(i)$ is the neighborhood set of $v_i$. Here, unlike many attention modules that normalizes the attention scores so that the summantion of the scores becomes 1.0, we normalize each attention score to be between 0.0 and 1.0. We empirically confirmed that this selection results in better prediction performance than the conventional attention scheme.

\paragraph{FuncGC.} FuncGC is designed to model the higher-order interactions that are defined by the chemically-valid functional groups. Although the same functional groups can be present in many molecules, the effects that they have on the molecular properties may differ depending on their neighboring functional groups (or atoms). To account for such differences, we utilize the updated node feature that contains local information from the molecular graphs when generating the localized functional group features. We start the FuncGC by updating $z_k$ using $x_i'$ as follows:
\begin{align}
    \label{eq:fg-level-node-to-he}
    \tilde z_k = g_{\text{atom} \rightarrow \text{fg}}\Big(z_k, \sum_{i \in \mathcal{H}_k}x_i'\Big)
\end{align}
where $\tilde z_k$ is the localized feature that receives localized information from AtomGC, and $g_{\text{atom} \rightarrow \text{fg}}(\cdot)$ is the localizing MLP. Unlike $\gG$, $\gH$ has no naturally defined edges as the functional groups are concepts rather than physically exist. Hence, the edges among the hyperedges are learnt as follows:
\begin{align}
    \label{eq:fg-level-he-to-he}
    z'_{km} = g_{\text{edge}}(\tilde z_k,  \tilde z_m)
\end{align}
where $z'_{km}$ is the learnt edge feature between $\mathcal{H}_k$ and $\mathcal{H}_m$, and $g_{\text{edge}}$ is the edge MLP. 
$z'_{km}$ thus captures the interaction between $\mathcal{H}_k$ and $\mathcal{H}_m$. Lastly, the updated hyperedge features $z_k'$ is computed with $z'_{km}$ as follows:
\begin{align}
    \label{eq:fg-level-he-update}
    \beta_{km} &= g_{\text{attn}}(\tilde z_k, \tilde z_m) \\
    z'_k &= g_{\text{fg}}\Big(z_k, \sum_{m \in \mathcal{H}} \beta_{km} z'_{km}\Big)
\end{align}
where $\beta_{km}$ is the attention coefficient between the $k^{\text{th}}$ and the $m^{\text{th}}$ hyperedge, $g_{\text{attn}}(\cdot)$ is the attention MLP whose output activation is sigmoid activation function as in AtomGC, and $g_{\text{fg}}(\cdot)$ is the hyperedge update function. The {\fontfamily{lmtt}\selectfont HyperMP} layer is then used to modify the membership of the hyperedges and predict the molecular properties of the molecules using their respective computed embeddings.

Note that we did not design a path that propagate $z'_k$ back to the members (atoms) of $\mathcal{H}_k$. This design works similar to the uninterrupted gradient path of LSTM \citep{hochreiter1997long} or the latent arrays of Perciever models \citep{jaegle2021perceiver}. We also experimentally confirmed that this design shows better prediction results.

\subsection{Modifying the prior knowledge-guided structures}
\label{subsection:learning_subgraph}
$\mathcal{H}$ is constructed using the functional groups of the molecules. However, it is difficult to determine which prior knowledge to exploit in practice, and the wrong selection of prior knowledge can deteriorate the performances of the model. Hence, we allow models to adjust the members of $\mathcal{H}_k$, which is built upon the prior knowledge of functional groups, while predicting the molecular property. The general equation of the membership adjustment function  $F_\theta(\gG, \gH)$ can be defined as:
\begin{align}
    F_\theta(\gG, \gH) &= \tilde \gH
\end{align}
where $\tilde \gH$ is the membership-adjusted hypergraph. It first uses the membership encoder $f_\theta(\cdot)$ to produce the membership-encoded features as follows:
\begin{align}
     \{\hat x_{i} \}, \{\hat z_{k} \} &= f_\theta(\gG, \gH)
\end{align}
where $\hat x_{i}$ and $\hat z_{k}$ are the membership-encoded node and hyperedge features respectively, and $f_\theta(\cdot)$ is a stack of the {\fontfamily{lmtt}\selectfont HyperMP} layer(s). As the memberships can be interpreted as a virtual ``edge" between an atom $v_i$ and its functional group $\gH_k$, we employ a graph structure learning method to modify the membership. In the adjustment procedure, we consider the random discrete methods (i.e., the adjusted memberships are binary) which share a common philosophy with the Bayesian approaches. The membership adjustment procedure then starts by using the membership-encoded features to produce $\tilde \gH$ as follows:
\begin{align}
    m_{ik} &= g_\theta(\hat x_i, \hat z_k) \quad &\forall v_i \in \mathcal{H}_k \\
    \tilde m_{ik} &= \text{sigmoid}\big(\big(\log\big(\frac{m_{ik}}{1 - m_{ik}}\big) + \epsilon_0 - \epsilon_1\big) /s\big) \quad &\forall v_i \in \mathcal{H}_k
\end{align}
where $m_{ik}$ is the bernoulli parameter that models the probability of the event that $v_i$ becomes a member of $\gH_k$, $\tilde m_{ik}$ is the sampled membership, $g_\theta(\cdot)$ is the MLP whose output activation is the sigmoid function, $\epsilon_0$ and $\epsilon_1$ are the samples of Gumbel(0,1), and $s>0$ is the temperature parameter. This procedure reparameterize the Bernoulli distribution via Gumbel reparameterization such that the (sampled) binary $\tilde m_{ik}$ are differentiable \citep{jang2016categorical}. By annealing $s \rightarrow 0$, we can recover $\tilde m_{ik} \sim \text{Ber}(m_{ik})$. We define
the $k^\text{th}$ adjusted hyperedge $\tilde \gH_k=\{v_i \in \gH_k \,|\, \tilde m_{ik} = 1\}$. $\tilde \gH$ will then be used to produce the final predictions.

A similar approach is investigated in the context of pair-wise graph structure learning \citep{shang2021discrete}, where they assume that the edges of a complete graph is subjective to edge learning. On the other hand, we utilize this idea only to the members of hyperedges so as to provide a balance between the usage of prior knowledge and the data-driven scheme.

\subsection{Molecular properties prediction}  
From the aforementioned methods, $\gH$ is first constructed using the prior knowledge of the functional groups from a given $\gG$. The memberships of $\gH$ are then adjusted using $F_\theta(\gG, \gH)$ to produce $\tilde \gH$. Hence, in the final step of {\fontfamily{lmtt}\selectfont MolHMPN}, the target label $y$ of a given molecule is predicted by updating $\gG$ and $\tilde \gH$ using the {\fontfamily{lmtt}\selectfont HyperMP} layer as follows:
\begin{align}
    y = G_\theta(\gG, \tilde \gH)
\end{align}
where $G_\theta(\cdot)$ is the property prediction function, which consists of a stack of the {\fontfamily{lmtt}\selectfont HyperMP} layer(s), a readout function, and a MLP.

\subsection{Training model}
For our tasks, we randomly split the datasets into 80:10:10 ratio as the training, validation and test sets and take the average of the results from different 5 random seeds (0 to 4).
The atom and bond features that are used as the initial node and edge features are given in Appendix \ref{appendix:training}. For $F_\theta(\cdot)$ and $G_\theta(\cdot)$, we use only one {\fontfamily{lmtt}\selectfont HyperMP} layer each. The attentive sum and max function are used as the readout function of $G_\theta(\cdot)$. The loss functions for the classification and regression tasks are the binary cross-entropy (BCE) and mean squared error (MSE), respectively. We give extra weights to the minority class in the loss functions for the classification datasets based on the ratio of the minority to majority class of each task to handle the class imbalance problems. We train {\fontfamily{lmtt}\selectfont MolHMPN} by minimizing the loss functions. We use the AdamP optimizer \citep{adamp}, whose learning rate is initialized as 0.001 and scheduled by the cosine annealing method \citep{consine}. We train the models for 500 epochs with mini-batch size of 512. More training details can be found in Appendix \ref{appendix:training}. The results at the end of the training are recorded.

\section{Results and discussion}
In this section, we evaluate the performance of {\fontfamily{lmtt}\selectfont MolHMPN} with several experiments. We first evaluate the performance of {\fontfamily{lmtt}\selectfont MolHMPN} with several baseline methods that consider the pair-wise and/or high-order interactions in the molecules in Section \ref{subsection:baseline_comp}. We then perform various ablation studies and show that it is beneficial to:
\vspace{-0.2cm}
\begin{itemize}[leftmargin=*]
    \item \textbf{Section \ref{subsection:design_abalation}}: account for both the pair-wise and higher-order connectivities.
    \item \textbf{Section \ref{subsection:subg_ablation}}: incorporate functional group information as the means of constructing hyperedge.
    \item \textbf{Section \ref{subsection:extension_ablation}}: leverage the prior knowledge \textit{partially} with the extended hyperedges. 
\end{itemize}

The datasets that are used for evaluation includes Tox21, ClinTox, SIDER, BBBP, BACE, ESOL, FreeSolv and Lipophilicity. The detailed dataset description can be found in Appendix \ref{appendix:training}. We provide the benchmark results and visualizations of {\fontfamily{lmtt}\selectfont MolHMPN} to evaluate the performances. 

\subsection{Comparing with baseline models}
\label{subsection:baseline_comp}
We evaluate the performance of {\fontfamily{lmtt}\selectfont MolHMPN} with baselines that make use of the pair-wise connectivities  (\textit{PAIR}) and/or higher-order connectivities (\textit{HIGH}) in the molecules. For the \textit{PAIR} baselines, we consider models using atoms only (MPNN (atom only)) \citep{dmpnn:2019}, atom and bonds (MPNN) \citep{dmpnn:2019}, directed bonds (DMPNN) \citep{dmpnn:2019}, and atoms and bonds with enhanced interactions (CMPNN) \citep{cmpnn:2020}. For the \textit{HIGH} baselines, we compare with baselines that have used substructures whose nodes are not connected by an edge (AGCN) \citep{agcn:2018}, substructures with marginal nodes (GAAN) \citep{gaan:2019}, and substructures that are constructed by junction tree (ML-MPNN) \citep{molkit:2021}. The results of the baselines are taken directly from their respective papers, except for CMPNN\footnote{We rerun their codes for all datasets as a mistake was found in their results as stated in their official code \textcolor{blue}{https://github.com/SY575/CMPNN.git}}.

\begin{table}[h]
\caption{\textbf{Benchmark results.} Comparing between different methods for molecular properties prediction. All results are taken from the original papers except CMPNN. Results in red are the best-performing results, and the results in blue are the second best-performing results. ($\uparrow$ means that higher result is better and $\downarrow$ means that lower result is better.)}
\vskip 0.05in
\resizebox{\columnwidth}{!}{
\begin{tabular}{ll|ccccc|ccc}
\toprule
& Metric & \multicolumn{5}{c|}{AUROC (Classification)}  & \multicolumn{3}{c}{RMSE (Regression)} \\
\hline
& Dataset
& Tox21 ($\uparrow$) 
& ClinTox ($\uparrow$)
& SIDER ($\uparrow$)
& BBBP ($\uparrow$)
& BACE ($\uparrow$) 
& ESOL ($\downarrow$) 
& FreeSolv ($\downarrow$)
& Lipophilicity ($\downarrow$) \\
\hline 
\textit{PAIR}
& \tikz\draw[repr,fill=repr](0,0) circle (.5ex); MPNN (atom only) 
& $0.845$ 
& $0.896$ 
& $0.644$ 
& $0.908$ 
& $0.864$ 
& $0.719$ 
& $1.243$ 
& $0.625$ \\
& \textcolor{repr}{$\star$} MPNN  
& $0.844$ 
& $0.881$ 
& $0.641$ 
& $0.910$ 
& $0.850$ 
& $0.702$ 
& $1.242$ 
& $0.645$ \\
& \textcolor{repr}{$\times$} DMPNN
& $0.845$ 
& $0.894$ 
& $0.646$ 
& $0.913$ 
& $0.878$ 
& $0.665$ 
& $1.167$ 
& $0.596$ \\
& \begin{tikzpicture}
\node[color=repr](0,0){\pgfuseplotmark{triangle*}};\end{tikzpicture} CMPNN
& \textcolor{red}{\textbf{0.854}} 
& \textcolor{blue}{\textbf{0.908}} 
& $0.656$ 
& \textcolor{red}{\textbf{0.958}}
& \textcolor{blue}{\textbf{0.887}}
& $0.567$ 
& \textcolor{blue}{\textbf{0.901}}
& $0.582$ \\
\hline
\textit{HIGH}
& \tikz\draw[sub,fill=sub](0,0) circle (.5ex); AGCN
& $0.802$ 
& $0.868$ 
& $0.592$ 
& $-$ 
& $-$ 
& \textcolor{blue}{\textbf{0.306}} 
& $1.33$
& $0.736$ \\
& \textcolor{sub}{$\star$} GAAN
& $0.839$ 
& $0.888$ 
& \textcolor{blue}{\textbf{0.658}} 
& $-$ 
& $-$ 
& \textcolor{red}{\textbf{0.294}}
& $1.057$ 
& $0.605$ \\
& \textcolor{sub}{$\times$} ML-MPNN
& \textcolor{blue}{\textbf{0.852}} 
& $0.892$ 
& \textcolor{red}{\textbf{0.689}} 
& $-$ 
& $-$ 
& $0.571$ 
& $1.052$ 
& \textcolor{blue}{\textbf{0.560}} \\
\hline
& \tikz\draw[molhgcn,fill=molhgcn](0,0) circle (.5ex); MolHMPN 
& $0.840$ 
& \textcolor{red}{\textbf{0.919}} 
& $0.617$ 
& \textcolor{blue}{\textbf{0.940}} 
& \textcolor{red}{\textbf{0.892}}
& 0.390
& \textcolor{red}{\textbf{0.813}}
& \textcolor{red}{\textbf{0.514}} \\
\bottomrule
\end{tabular}
}
\label{tab:benchmark}
\end{table}

\begin{figure}
    \centering
    \begin{subfigure}[b]{0.8\textwidth}
    \includegraphics[width={\linewidth}]{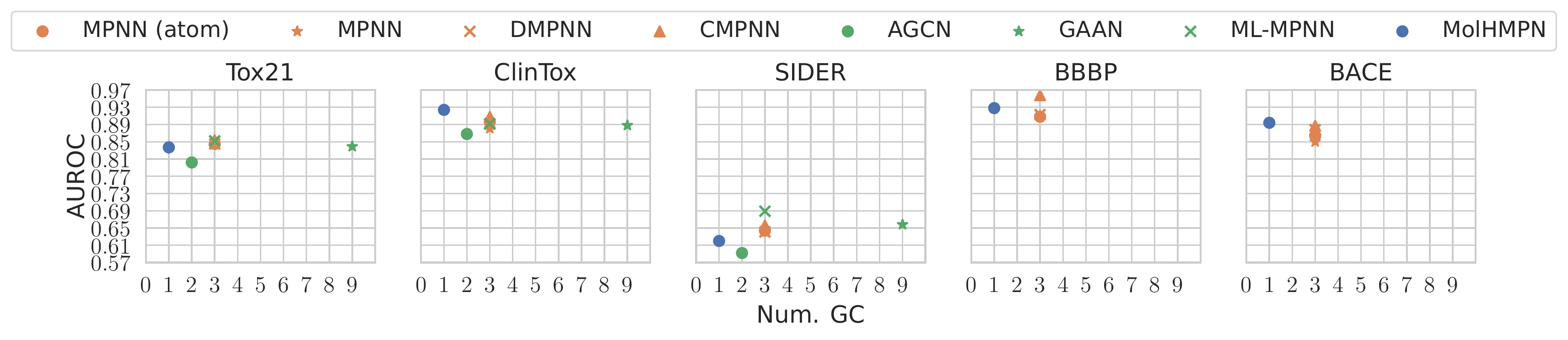}
    \end{subfigure}
    \begin{subfigure}[b]{0.8\textwidth}
    \includegraphics[width={\linewidth}]{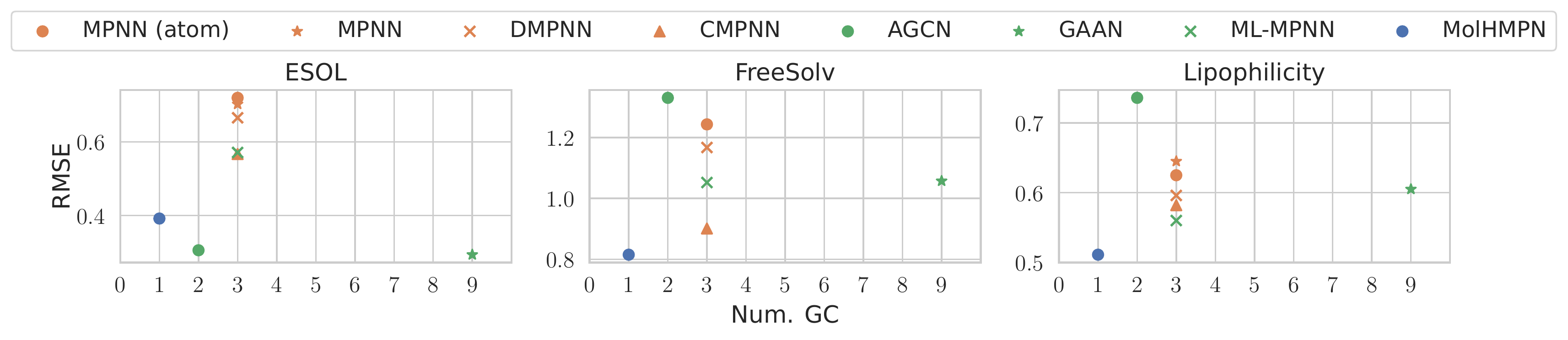}
    \end{subfigure}
    \caption{\textbf{Number of graph convolutions vs. classification performances}}
    \label{fig:ngc_perf}
\end{figure}

Table \ref{tab:benchmark} shows the overall results of {\fontfamily{lmtt}\selectfont MolHMPN} on graph classification and regression tasks. From the results, we can see that {\fontfamily{lmtt}\selectfont MolHMPN} outperforms the other baselines for four out of eight datasets. 
From the \textit{PAIR} results, we can see that the usage of atoms, undirected and directed bond information do not have a significant impact on the performances. Instead, increasing the interactions between the atoms and bonds (CMPNN) gives better results, especially for BBBP, ESOL and FreeSolv. Comparing {\fontfamily{lmtt}\selectfont MolHMPN} with the \textit{PAIR} models, we can see that the inclusion of higher-order connectivities is indeed beneficial for the tasks as 
{\fontfamily{lmtt}\selectfont MolHMPN} outperforms the models for five out of eight datasets. Although the \textit{PAIR} models can capture higher-order connectivities by using multiple layers, {\fontfamily{lmtt}\selectfont MolHMPN} has outperformed the baselines with only one {\fontfamily{lmtt}\selectfont HyperMP} layer as shown in \Figref{fig:ngc_perf}. From the \textit{HIGH} results, we can see that {\fontfamily{lmtt}\selectfont MolHMPN} outperforms the other baselines for three out of six datasets. Compared to the other \textit{HIGH} models, ML-MPNN is the most similar to {\fontfamily{lmtt}\selectfont MolHMPN} as it integrates information from the nodes, edges, subgraphs and graphs, while {\fontfamily{lmtt}\selectfont MolHMPN} integrates information from the nodes, edges and subgraphs. However, {\fontfamily{lmtt}\selectfont MolHMPN} has outperformed ML-MPNN for four out of six datasets. This shows the efficacy of employing prior knowledge-guided hyperedges when conducting the benchmark tasks. Overall, the results shows the efficacy of accounting for both the pair-wise and higher-order connectivities, and employing prior knowledge-guided hyperedges.

\begin{figure}
    \centering
    \begin{subfigure}[b]{0.32\textwidth}
    \includegraphics[width={\linewidth}]{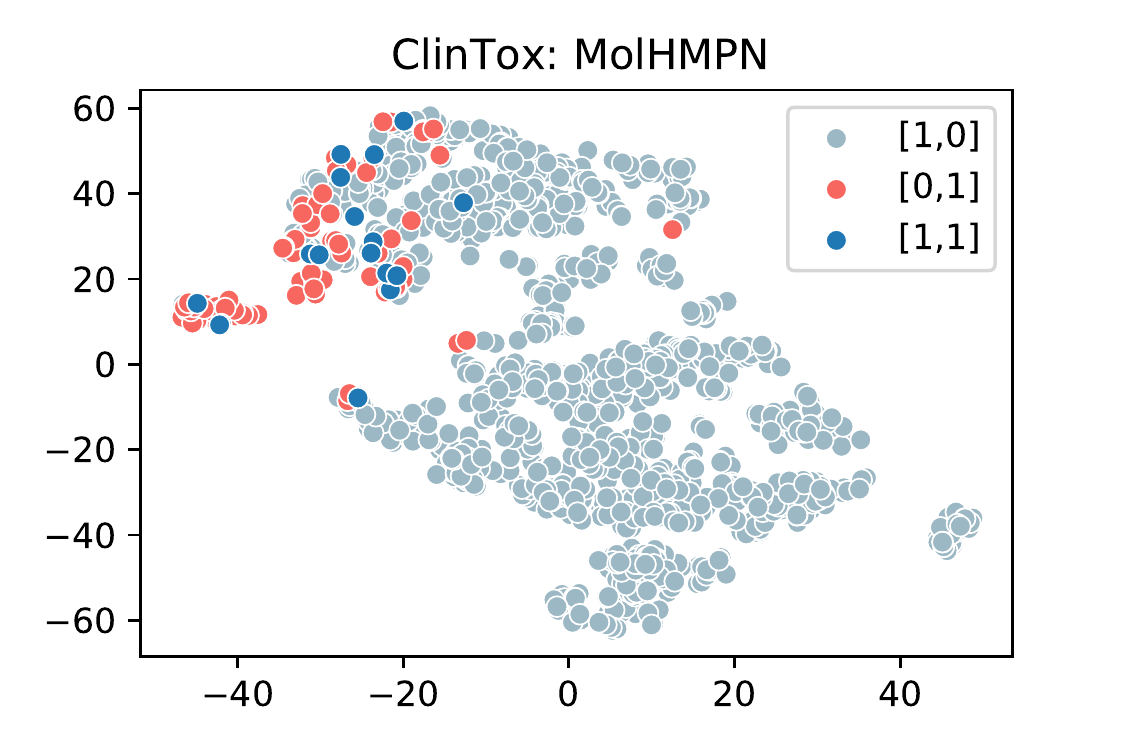}
    \end{subfigure}
    \begin{subfigure}[b]{0.32\textwidth}
    \includegraphics[width={\linewidth}]{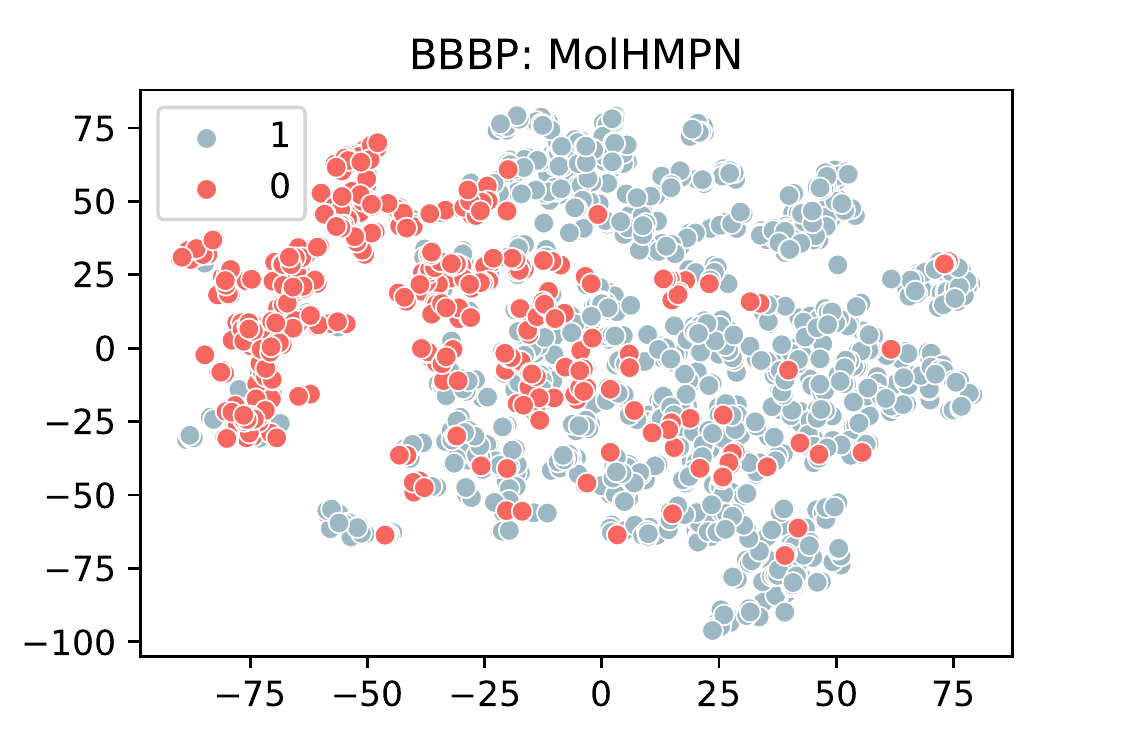}
    \end{subfigure}
    \begin{subfigure}[b]{0.32\textwidth}
    \includegraphics[width={\linewidth}]{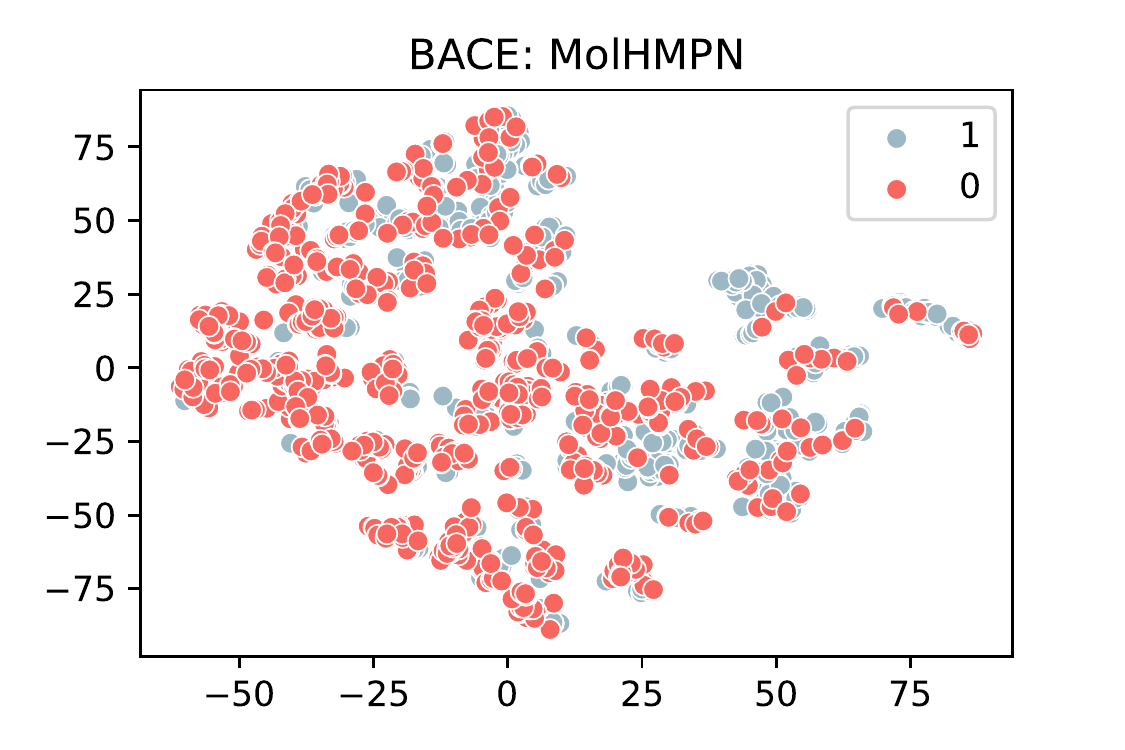}
    \end{subfigure}
    \begin{subfigure}[b]{0.32\textwidth}
    \includegraphics[width={\linewidth}]{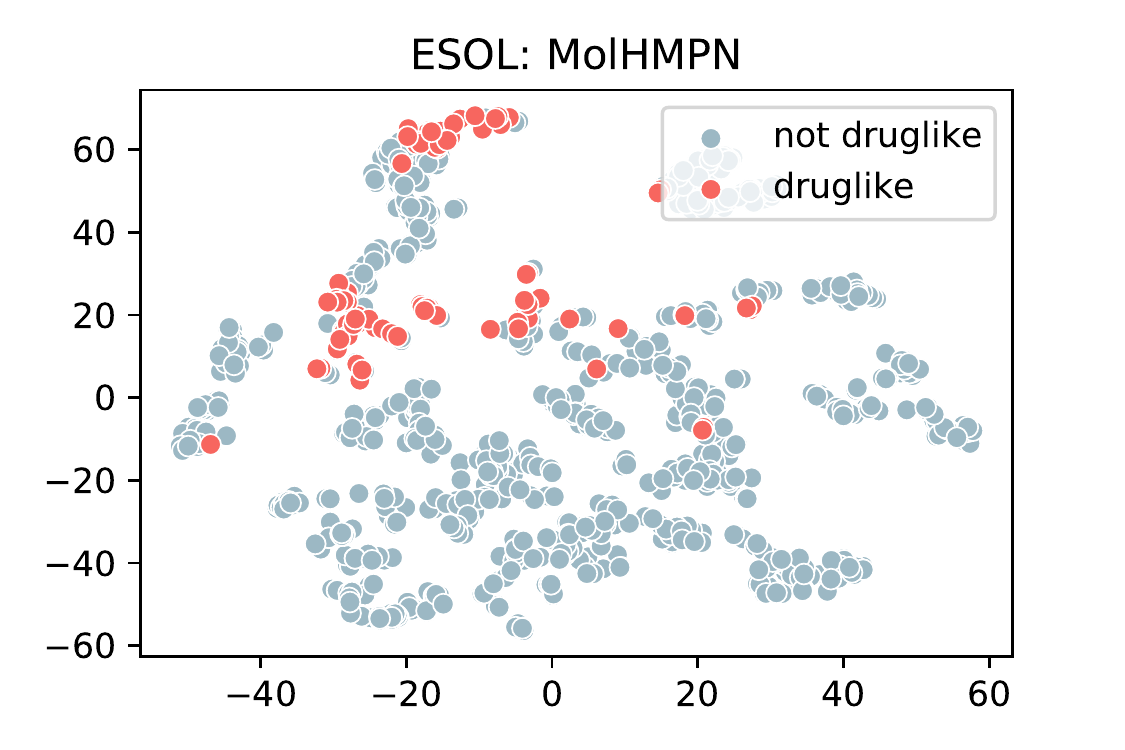}
    \end{subfigure}
    \begin{subfigure}[b]{0.32\textwidth}
    \includegraphics[width={\linewidth}]{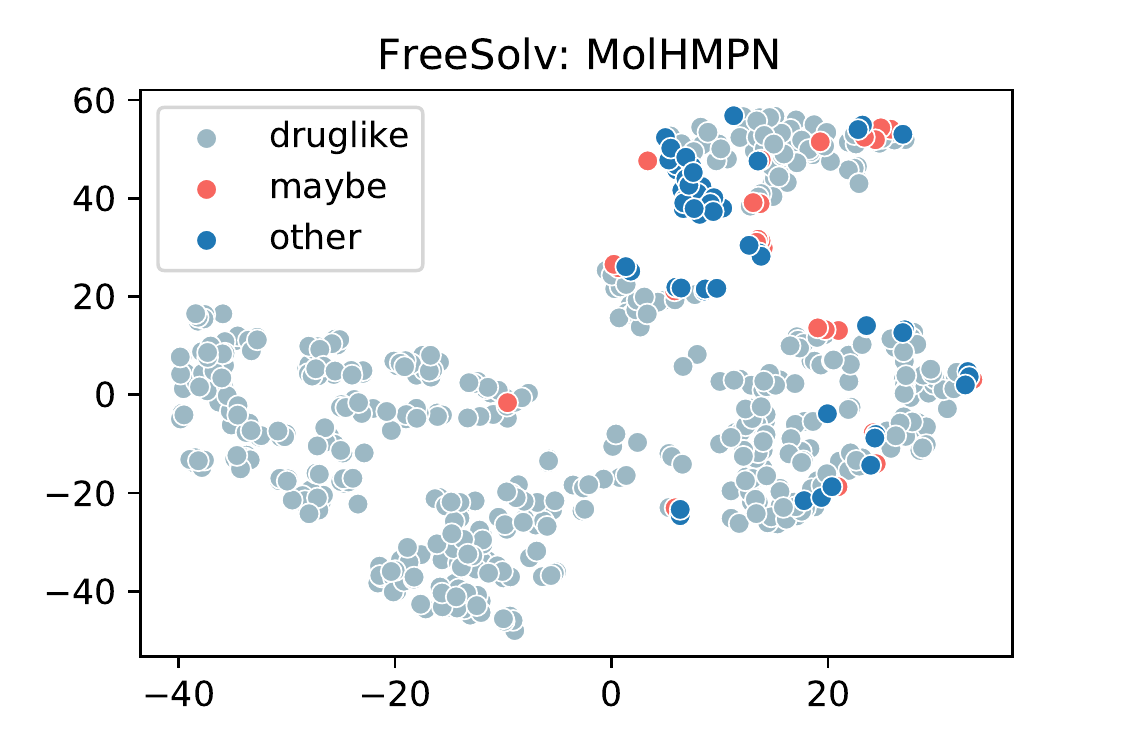}
    \end{subfigure}
    \begin{subfigure}[b]{0.32\textwidth}
    \includegraphics[width={\linewidth}]{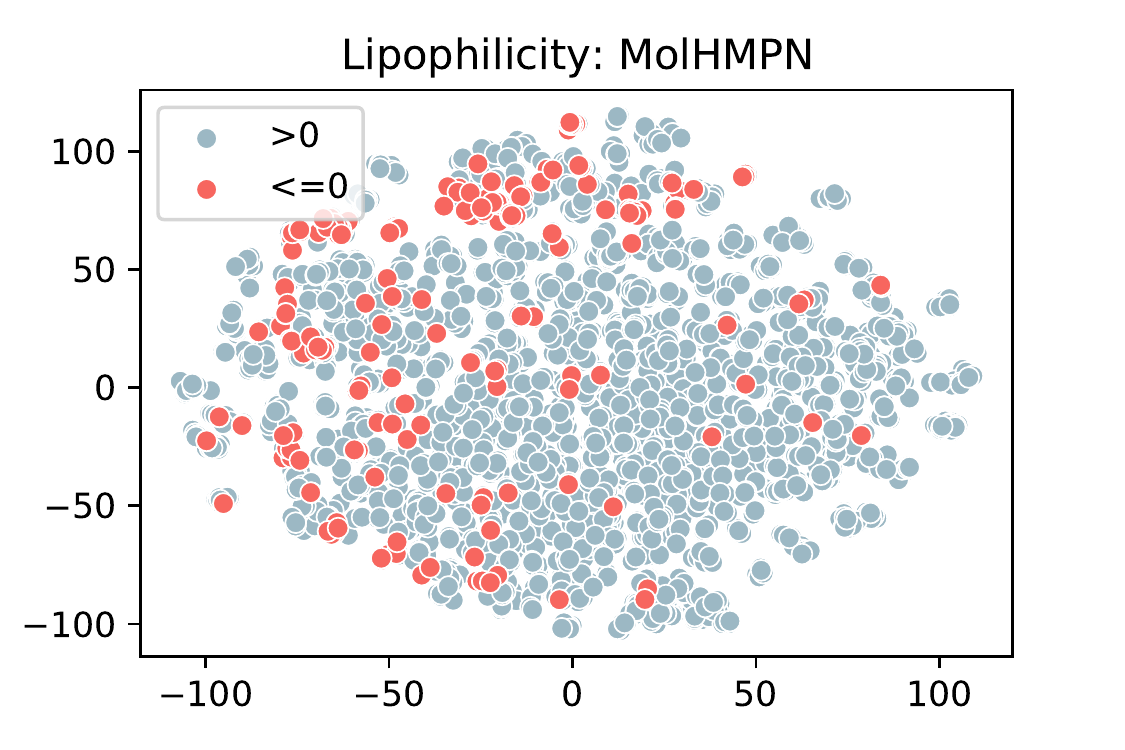}
    \end{subfigure}
    \caption{\textbf{{\fontfamily{lmtt}\selectfont MolHMPN} t-SNE plots}}
    \label{fig:baseline_tsne}
\end{figure}

\Figref{fig:baseline_tsne} shows the t-distributed stochastic neighbor embedding (t-SNE) plots of {\fontfamily{lmtt}\selectfont MolHMPN} for all of the datasets, except for Tox21 and SIDER since it is not straightforward to define the toxicity or druglikeliness of the molecules from the data labels. The classifications for ClinTox, BACE and BBBP are straightforward since they have less than 3 binary classification tasks. For the ESOL dataset, the molecules are considered to be druglike is their $logP$ value is between -0.4 to 5.6 \cite{logp}. For Freesolv, the molecules are considered to be druglike if their hydration free energy is below 0 kcal/mol, 0 to 1 kcal/mol for maybe druglike, and more than 1 kcal/mol for others \cite{hydration}. For Lipophilicity, the molecules are considered to be druglike if their $logD_{7.4}$ is above 0 \cite{logd}. The t-SNE plots will be analyzed in detail in the ablation studies. 

\subsection{Contribution of the pair-wise and higher-order connectivities}
\label{subsection:design_abalation}
In Section 1, we made the hypothesis that both the pair-wise and higher-order connectivities should be accounted for in molecular properties predictions when we compare the structure of xanthine, theobromine and pentoxifylline. In the {\fontfamily{lmtt}\selectfont HyperMP} layer, AtomGC accounts for the pair-wise connectivities while FuncGC accounts for the higher-order connectivities. Hence, to test this hypothesis, we devise the following variants:

\begin{itemize}[leftmargin=*]
    \item {\fontfamily{lmtt}\selectfont AtomGC}: the entire FuncGC in the {\fontfamily{lmtt}\selectfont HyperMP} layer is removed. Only the updated node features in the {\fontfamily{lmtt}\selectfont HyperMP} layer are used for prediction. 
    \item {\fontfamily{lmtt}\selectfont FuncGC}: the entire AtomGC in the {\fontfamily{lmtt}\selectfont HyperMP} layer is removed (i.e. the hyperedge features are no longer localized by the node features as in Equation \ref{eq:fg-level-node-to-he}. The localized features in Equations \ref{eq:fg-level-node-to-he}, \ref{eq:fg-level-he-to-he} and \ref{eq:fg-level-he-update} are the mean of the features of the nodes that are in the hyperedges). Only updated hyperedge features in the {\fontfamily{lmtt}\selectfont HyperMP} layer are used for prediction.
    \item {\fontfamily{lmtt}\selectfont MolHMPN-NoMod}: {\fontfamily{lmtt}\selectfont MolHMPN} with the original functional group design; the hyperedge extension and modificaton schemes are not implemented.
\end{itemize}

We evaluate the benchmark results and t-SNE plots of these variants. Note that the evaluation is done mainly for {\fontfamily{lmtt}\selectfont AtomGC}, {\fontfamily{lmtt}\selectfont FuncGC} and {\fontfamily{lmtt}\selectfont MolHMPN-NoMod} since we are focusing on the analysis of the the pair-wise and higher-order connectivities contribution to the prediction performance, and not the hyperedge extension and modification in this section. 

\begin{table}[h]
\centering
\caption{\textbf{Design analysis.} Comparison between {\fontfamily{lmtt}\selectfont AtomGC}, {\fontfamily{lmtt}\selectfont FuncGC} and {\fontfamily{lmtt}\selectfont MolHMPN-NoMod}. Results in bold are the best-performing results. ($\uparrow$ means that higher result is better and $\downarrow$ means that lower result is better.)}
\resizebox{\columnwidth}{!}{
\begin{tabular}{l|ccccc|ccc}
\toprule 
Metric & \multicolumn{5}{c|}{AUROC} & \multicolumn{3}{c}{RMSE} \\
\hline
Dataset
& Tox21 ($\uparrow$) 
& ClinTox ($\uparrow$)
& SIDER ($\uparrow$)
& BBBP ($\uparrow$)
& BACE ($\uparrow$) 
& ESOL ($\downarrow$) 
& FreeSolv ($\downarrow$)
& Lipophilicity ($\downarrow$) \\
\hline 
{\fontfamily{lmtt}\selectfont AtomGC}
& \begin{tabular}[c]{@{}c@{}} 0.832 \\ ($\pm$ 0.0175) \end{tabular}
& \begin{tabular}[c]{@{}c@{}} 0.897 \\ ($\pm$ 0.0390) \end{tabular}
& \begin{tabular}[c]{@{}c@{}} \textbf{0.606} \\ ($\pm$ 0.0.01575)
\end{tabular}
& \begin{tabular}[c]{@{}c@{}} 0.901 \\ ($\pm$ 0.0197)
\end{tabular}
& \begin{tabular}[c]{@{}c@{}} 0.869 \\ ($\pm$ 0.0269) \end{tabular}
& \begin{tabular}[c]{@{}c@{}}  \textbf{0.436} \\ ($\pm$ 0.0552) \end{tabular}
& \begin{tabular}[c]{@{}c@{}} 1.007 \\ ($\pm$ 0.3689)
\end{tabular}
& \begin{tabular}[c]{@{}c@{}} 0.561 \\ ($\pm$ 0.0786) \end{tabular}\\ 
{\fontfamily{lmtt}\selectfont FuncGC}
& \begin{tabular}[c]{@{}c@{}} 0.818 \\ ($\pm$ 0.0122) \end{tabular}
& \begin{tabular}[c]{@{}c@{}} 0.830 \\ ($\pm$ 0.05641) \end{tabular}
& \begin{tabular}[c]{@{}c@{}} 0.584 \\ ($\pm$ 0.0307) \end{tabular}
& \begin{tabular}[c]{@{}c@{}} 0.887 \\ ($\pm$ 0.0303) \end{tabular}
& \begin{tabular}[c]{@{}c@{}} 0.822 \\ ($\pm$ 0.0173) \end{tabular}
& \begin{tabular}[c]{@{}c@{}} 0.668 \\ ($\pm$ 0.1163) \end{tabular}
& \begin{tabular}[c]{@{}c@{}} 1.678 \\ ($\pm$ 0.6344) \end{tabular}
& \begin{tabular}[c]{@{}c@{}} 0.782 \\ ($\pm$ 0.0502) \end{tabular}\\
\toprule
{\fontfamily{lmtt}\selectfont MolHMPN-NoMod}
& \begin{tabular}[c]{@{}c@{}} \textbf{0.839} \\ ($\pm$ 0.0147) \end{tabular}
& \begin{tabular}[c]{@{}c@{}} \textbf{0.909} \\ ($\pm$ 0.0394) \end{tabular}
& \begin{tabular}[c]{@{}c@{}} 0.605 \\ ($\pm$ 0.0227) \end{tabular}
& \begin{tabular}[c]{@{}c@{}} \textbf{0.928} \\ ($\pm$ 0.0302) \end{tabular}
& \begin{tabular}[c]{@{}c@{}} \textbf{0.892} \\ ($\pm$ 0.0232) \end{tabular}
& \begin{tabular}[c]{@{}c@{}} 0.447 \\ ($\pm$ 0.0559) \end{tabular}
& \begin{tabular}[c]{@{}c@{}} \textbf{0.813} \\ ($\pm$ 0.3952) \end{tabular}
& \begin{tabular}[c]{@{}c@{}} \textbf{0.517} \\ ($\pm$ 0.0415) \end{tabular}\\ 
\bottomrule
\end{tabular}
}
\label{tab:atomgcfuncgc}
\end{table}

Table \ref{tab:atomgcfuncgc} shows the results of {\fontfamily{lmtt}\selectfont AtomcGC}, {\fontfamily{lmtt}\selectfont FuncGC} and {\fontfamily{lmtt}\selectfont MolHMPN-NoMod}. From Table \ref{tab:atomgcfuncgc}, we can see that {\fontfamily{lmtt}\selectfont MolHMPN-NoMod} outperforms the other models for six out of eight of the datasets, and more significantly for FreeSolv. We conjecture that this is because FreeSolv contain fragment-like compounds where the typical size of the molecules is substantially smaller than typical small-molecule drugs \cite{mobley2014freesolv}. This makes the molecules more similar to the hyperedge design and hence, contributing significantly to the performance of {\fontfamily{lmtt}\selectfont MolHMPN-NoMod} to FreeSolv. Among the three models, we can see that {\fontfamily{lmtt}\selectfont FuncGC} performs the worst. This may be because there is no naturally defined edges between each hyperedge since two hyperedges may have more than one of the same node, making it difficult to design an edge that can define the relationship between two hyperedge well. Nonetheless, this does not mean that the incorporation of the functional group information degrades the performance of the model. This can be verified by comparing the results of {\fontfamily{lmtt}\selectfont AtomGC} and {\fontfamily{lmtt}\selectfont MolHMPN-NoMod}. From the results, we can see that the incorporation of functional group information has increased the performance of {\fontfamily{lmtt}\selectfont MolHMPN-NoMod} as it has outperformed {\fontfamily{lmtt}\selectfont AtomGC} for six out of eight datasets. This shows that using both the pair-wise and higher-order connectivities information is beneficial to the performances. We further verify this by visualizing the distribution of readout values of each model via t-SNE.

\begin{figure}
    \centering
    \begin{subfigure}[b]{1.1\textwidth}
    \includegraphics[width={\linewidth}]{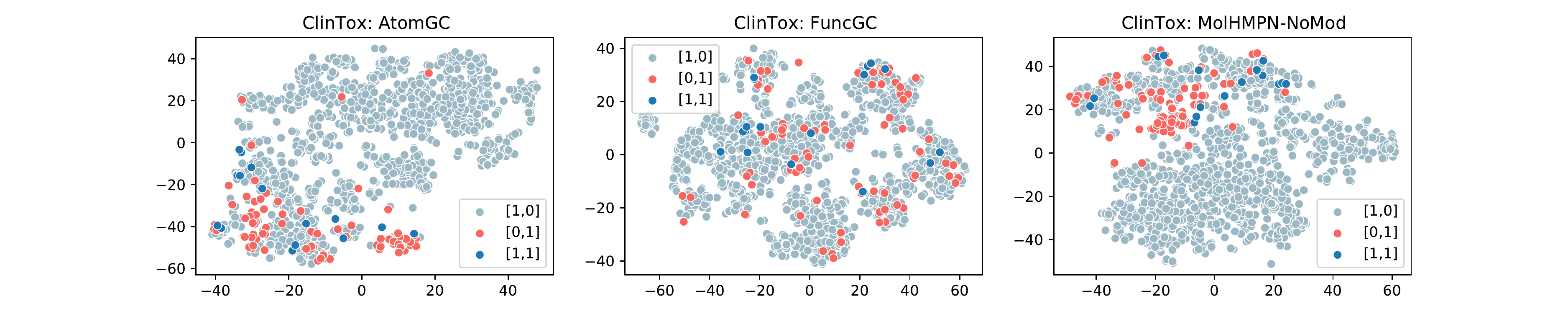}
    \end{subfigure}
    \begin{subfigure}[b]{1.1\textwidth}
    \includegraphics[width={\linewidth}]{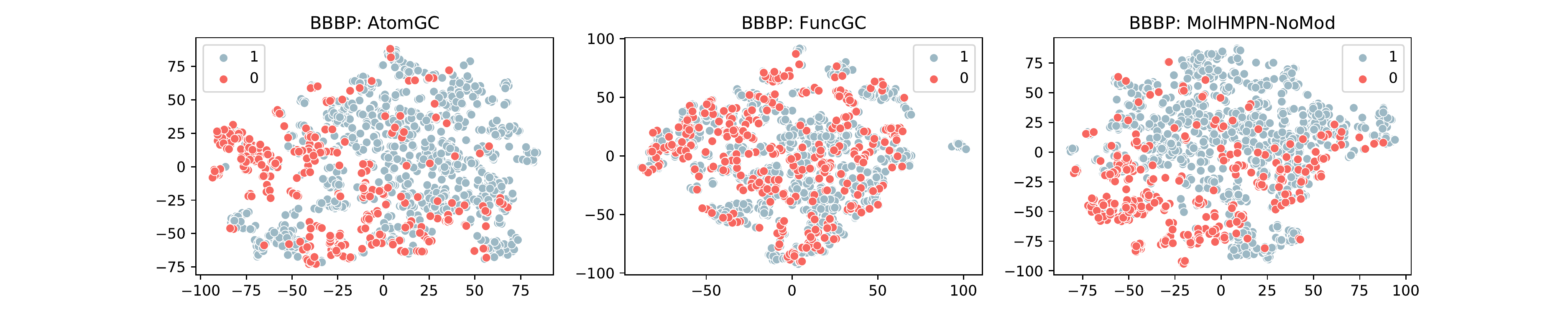}
    \end{subfigure}
    \begin{subfigure}[b]{1.1\textwidth}
    \includegraphics[width={\linewidth}]{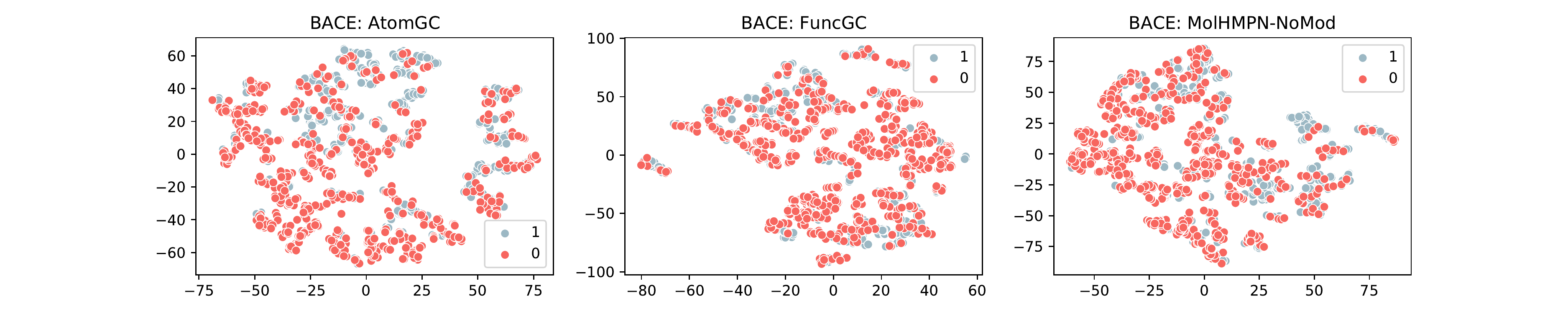}
    \end{subfigure}
    \begin{subfigure}[b]{1.1\textwidth}
    \includegraphics[width={\linewidth}]{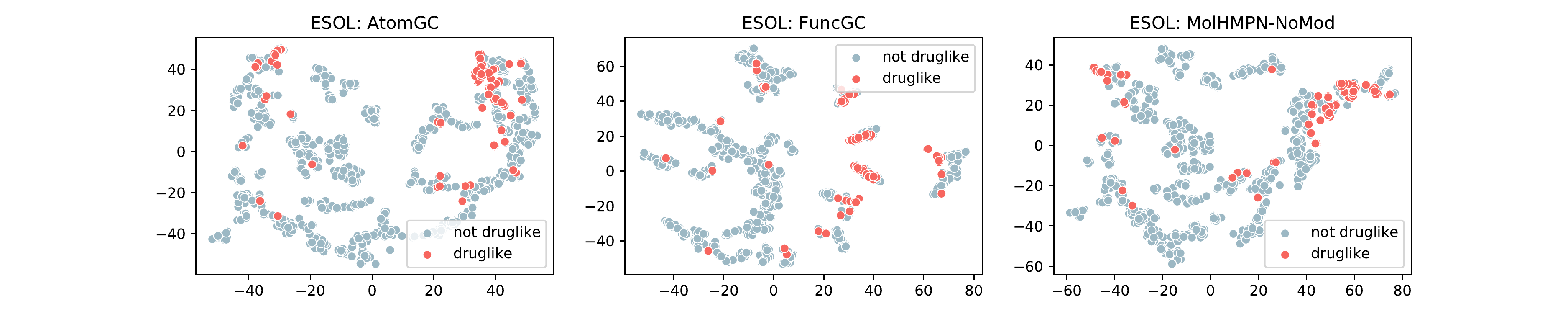}
    \end{subfigure}
    \begin{subfigure}[b]{1.1\textwidth}
    \includegraphics[width={\linewidth}]{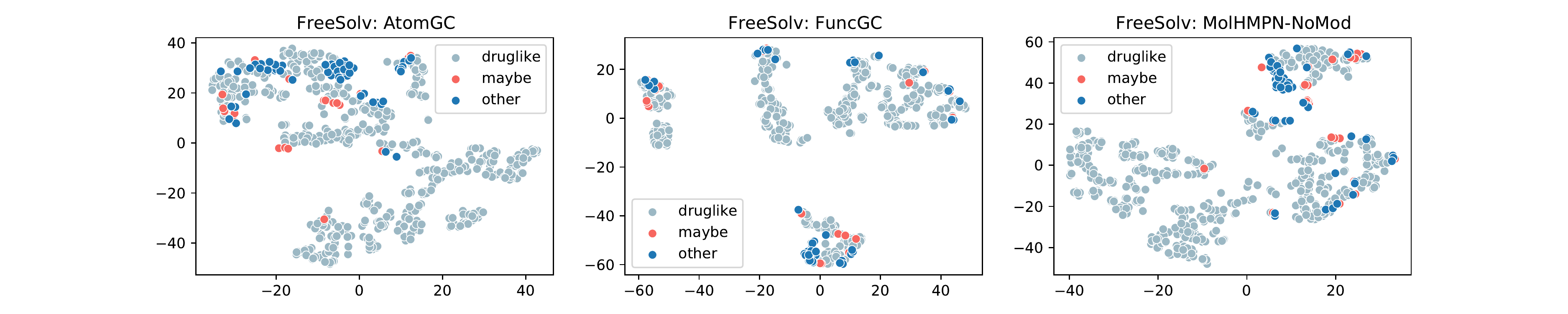}
    \end{subfigure}
    \begin{subfigure}[b]{1.1\textwidth}
    \includegraphics[width={\linewidth}]{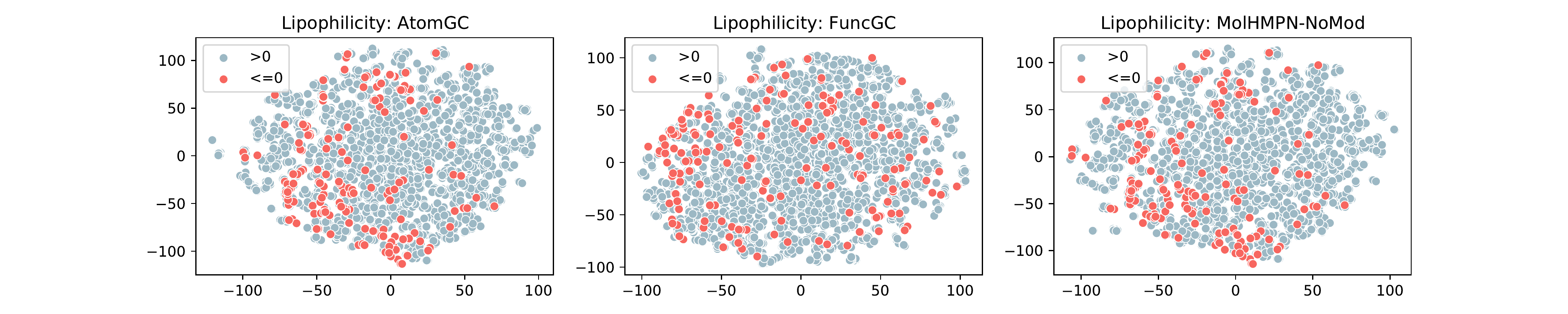}
    \end{subfigure}
    \caption{\textbf{t-SNE plots for {\fontfamily{lmtt}\selectfont AtomGC}, {\fontfamily{lmtt}\selectfont FuncGC} and {\fontfamily{lmtt}\selectfont MolHMPN-NoMod}.} Analyzing the contributions of pair-wise and/or higher-order connectivities in {\fontfamily{lmtt}\selectfont MolHMPN}}
    \label{fig:all_tsne}
\end{figure}


Figure \ref{fig:all_tsne} shows the t-SNE plot for the datasets. From the BACE and Lipophilicity plots, no clear separation can be observed. From the ESOL plots, it can be seen that {\fontfamily{lmtt}\selectfont AtomGC} and {\fontfamily{lmtt}\selectfont MolHMPN-NoMod} has similar data separation, which is consistent with the results in Table \ref{tab:atomgcfuncgc}. From the ClinTox, BBBP and FreeSolv plots, a clearer separation can be seen as the minority plots are closer to each other with the incorporation of both the atom and functional group information, which is also consistent with the results in Table \ref{tab:atomgcfuncgc} and our hypothesis that both the pair-wise (atom) and higher-order connectivities (functional groups) are important in molecular properties prediction tasks. This supports the benchmark results in Table \ref{tab:atomgcfuncgc} that integrating both the pair-wise and higher-order connectivities information is beneficial. 

Overall, from the results in Table \ref{tab:atomgcfuncgc} and the visualizations in Figure \ref{fig:all_tsne}, we can verify that incorporating both the pair-wise and higher-order connectivity information is beneficial to the performance of the model.

\subsection{Comparison with different subgraphs}
\label{subsection:subg_ablation}
In Section 1, we made the hypothesis that  substructures that are significant to the molecular properties should be used, and described the effects that functional groups have on the molecules. To test this hypothesis, we compare the performance of {\fontfamily{lmtt}\selectfont MolHMPN} with other methods that employ other kinds of subgraphs and are known to be effective in solving molecule generation and graph meta-learning tasks. Since we are only analyzing the effectiveness of functional groups in molecular properties prediction tasks, we use the results of {\fontfamily{lmtt}\selectfont MolHMPN-NoMod}. The subgraphs evaluated include:

\begin{itemize}[leftmargin=*]
    \item \textbf{Ring $\&$ Chemical Bond}: utilizes the set of ring structures and chemical bonds as the subgraph\footnote{In original paper, the frequently-occurring chemical subgraphs are also considered. However, in our benchmark datasets, none of the dataset satisfies the proposed value for occurrence frequency.} \citep{wgj:2020}
    \item \textbf{$K$-hop neighbors}: utilizes the $K$-hop neighbors as subgraphs \citep{gmeta:2020}
    \item \textbf{MolSoft}: utilizes the attention mechanism \citep{vaswani2017attention} to construct \textit{soft} subgraphs. Refer to Appendix \ref{appendix:ablation} for more details.
\end{itemize}

1-hop neighbor is an edge-centric reformulation of the original graph where each hyperedge contains two nodes that are connected together by an edge, and MolSoft is approach that does not depend on prior knowledge and attempts to capture substructures that have significant effects to the molecular properties. Hence, the subgraphs that does not necessarily depend on the prior knowledge in chemistry (\textbf{Ring $\&$ Chemical Bond} and \textbf{$K$-hop neighbors}) and subgraphs that does not depend on prior knowledge (\textbf{MolSoft}) are evaluated to measure the significance of functional group to molecular properties prediction.

In the experiments, we replace the hyperedge construction rules with those of the baseline methods, and assess their performances with our benchmark datasets. Other than the hyperedge construction method, we use the same model and experiment setups as in Appendix \ref{appendix:training} to make fair comparisons.

\begin{table}[h]
\caption{\textbf{Subgraph Comparison.} Comparison between different types of subgraphs. Results in red are the best-performing results, and the results in blue are the second best-performing results.  ($\uparrow$ means that higher result is better and $\downarrow$ means that lower result is better.)}
\resizebox{\columnwidth}{!}{
\begin{tabular}{c|ccccc|ccc}
\toprule 
Metric & \multicolumn{5}{c|}{AUROC} & \multicolumn{3}{c}{RMSE} \\
\hline
Dataset & 
Tox21 ($\uparrow$) &
ClinTox ($\uparrow$) &
SIDER ($\uparrow$) &
BBBP ($\uparrow$) &
BACE ($\uparrow$) &
ESOL ($\downarrow$) &
FreeSolv ($\downarrow$) & 
Lipophilicity ($\downarrow$) \\
\hline 
Ring $\&$ C. Bond & 
\begin{tabular}[c]{@{}c@{}} 0.834 \\ ($\pm$ 0.0142) \end{tabular} & 
\begin{tabular}[c]{@{}c@{}} \textcolor{blue}{\textbf{0.911}}  \\ ($\pm$ 0.0401) \end{tabular} & \begin{tabular}[c]{@{}c@{}} 0.577 \\ ($\pm$ 0.0339) \end{tabular} &
\begin{tabular}[c]{@{}c@{}} 0.919 \\ ($\pm$ 0.0124) \end{tabular} &
\begin{tabular}[c]{@{}c@{}} 0.884 \\ ($\pm$ 0.0106) \end{tabular} & 
\begin{tabular}[c]{@{}c@{}} 0.477  \\ ($\pm$ 0.0547) \end{tabular} & \begin{tabular}[c]{@{}c@{}} 1.468  \\ ($\pm$ 0.5970) \end{tabular} & \begin{tabular}[c]{@{}c@{}} 0.520 \\ ($\pm$ 0.0475) \end{tabular}\\
1-hop ngh.
& \begin{tabular}[c]{@{}c@{}} 0.833 \\ ($\pm$ 0.0194) \end{tabular}
& \begin{tabular}[c]{@{}c@{}} 0.895 \\ ($\pm$ 0.0273) \end{tabular}
& \begin{tabular}[c]{@{}c@{}} \textcolor{blue}{\textbf{0.597}} \\ ($\pm$ 0.0246) \end{tabular}
& \begin{tabular}[c]{@{}c@{}} 0.906 \\ ($\pm$ 0.0195) \end{tabular}
& \begin{tabular}[c]{@{}c@{}} 0.869 \\ ($\pm$ 0.0297) \end{tabular}
& \begin{tabular}[c]{@{}c@{}} 0.453 \\ ($\pm$ 0.0588) \end{tabular}
& \begin{tabular}[c]{@{}c@{}} 1.128 \\ ($\pm$ 0.3030) \end{tabular}
& \begin{tabular}[c]{@{}c@{}} 0.542 \\ ($\pm$ 0.1068) \end{tabular}\\ 
2-hop ngh.
& \begin{tabular}[c]{@{}c@{}} \textcolor{blue}{\textbf{0.836}} \\ ($\pm$ 0.0135) \end{tabular}
& \begin{tabular}[c]{@{}c@{}} 0.910 \\ ($\pm$ 0.0448) \end{tabular}
& \begin{tabular}[c]{@{}c@{}} 0.582 \\ ($\pm$ 0.0271) \end{tabular}
& \begin{tabular}[c]{@{}c@{}} \textcolor{blue}{\textbf{0.926}} \\ ($\pm$ 0.0314) \end{tabular}
& \begin{tabular}[c]{@{}c@{}} \textcolor{red}{\textbf{0.894}} \\ ($\pm$ 0.0240) \end{tabular}
& \begin{tabular}[c]{@{}c@{}} \textcolor{blue}{\textbf{0.431}} \\ ($\pm$ 0.0709) \end{tabular}
& \begin{tabular}[c]{@{}c@{}} \textcolor{blue}{\textbf{0.995}} \\ ($\pm$ 0.3844) \end{tabular}
& \begin{tabular}[c]{@{}c@{}} 0.564 \\ ($\pm$ 0.0475) \end{tabular}\\
3-hop ngh.
& \begin{tabular}[c]{@{}c@{}} 0.830 \\ ($\pm$ 0.0147) \end{tabular}
& \begin{tabular}[c]{@{}c@{}} 0.881 \\ ($\pm$ 0.0363) \end{tabular}
& \begin{tabular}[c]{@{}c@{}} 0.597 \\ ($\pm$ 0.0307) \end{tabular}
& \begin{tabular}[c]{@{}c@{}} 0.918 \\ ($\pm$ 0.0278) \end{tabular}
& \begin{tabular}[c]{@{}c@{}} 0.871 \\ ($\pm$ 0.0136) \end{tabular}
& \begin{tabular}[c]{@{}c@{}} 0.480 \\ ($\pm$ 0.1032) \end{tabular}
& \begin{tabular}[c]{@{}c@{}} 1.078 \\ ($\pm$ 0.3094) \end{tabular}
& \begin{tabular}[c]{@{}c@{}} 0.571 \\ ($\pm$ 0.0889) \end{tabular} \\
\hline
MolSoft
& \begin{tabular}[c]{@{}c@{}} 0.803 \\ ($\pm$ 0.0161) \end{tabular}
& \begin{tabular}[c]{@{}c@{}} \textcolor{red}{\textbf{0.915}} \\ ($\pm$ 0.0352) \end{tabular}
& \begin{tabular}[c]{@{}c@{}} 0.588 \\ ($\pm$ 0.0300) \end{tabular}
& \begin{tabular}[c]{@{}c@{}} 0.888 \\ ($\pm$ 0.0210) \end{tabular}
& \begin{tabular}[c]{@{}c@{}} 0.871 \\ ($\pm$ 0.0311) \end{tabular}
& \begin{tabular}[c]{@{}c@{}} \textcolor{red}{\textbf{0.420}} \\ ($\pm$ 0.0653) \end{tabular}
& \begin{tabular}[c]{@{}c@{}} 1.049 \\ ($\pm$ 0.4940) \end{tabular}
& \begin{tabular}[c]{@{}c@{}} \textcolor{blue}{\textbf{0.519}} \\ ($\pm$ 0.0535) \end{tabular}\\ 
\bottomrule
{\fontfamily{lmtt}\selectfont MolHMPN-NoMod}
& \begin{tabular}[c]{@{}c@{}} \textcolor{red}{\textbf{0.839}} \\ ($\pm$ 0.0147) \end{tabular}
& \begin{tabular}[c]{@{}c@{}} 0.909 \\ ($\pm$ 0.0394) \end{tabular}
& \begin{tabular}[c]{@{}c@{}} \textcolor{red}{\textbf{0.605}} \\ ($\pm$ 0.0227) \end{tabular}
& \begin{tabular}[c]{@{}c@{}} \textcolor{red}{\textbf{0.928}} \\ ($\pm$ 0.0302) \end{tabular}
& \begin{tabular}[c]{@{}c@{}} \textcolor{blue}{\textbf{0.892}} \\ ($\pm$ 0.0232) \end{tabular}
& \begin{tabular}[c]{@{}c@{}} 0.447 \\ ($\pm$ 0.0559) \end{tabular}
& \begin{tabular}[c]{@{}c@{}} \textcolor{red}{\textbf{0.813}} \\ ($\pm$ 0.3952) \end{tabular}
& \begin{tabular}[c]{@{}c@{}} \textcolor{red}{\textbf{0.517}} \\ ($\pm$ 0.0415) \end{tabular}\\ 
\bottomrule
\end{tabular}
}
\label{tab:ablacyc}
\end{table}

Table \ref{tab:ablacyc} shows the results where different types of subgraphs are used. From Table \ref{tab:ablacyc}, {\fontfamily{lmtt}\selectfont MolHMPN-NoMod} has outperformed the other methods for five out of eight datasets, especially for FreeSolv. For SIDER, {\fontfamily{lmtt}\selectfont MolHMPN-NoMod} has outperformed the other methods and has the smallest standard deviation. For BBBP, although 2-hop neighbor is comparable with {\fontfamily{lmtt}\selectfont MolHMPN-NoMod}, {\fontfamily{lmtt}\selectfont MolHMPN-NoMod} has a smaller standard deviation. One notable trend is that the 1-hop and 3-hop neighbor underperform as compared to 2-hop neighbor even though they model pair-wise and higher-order connectivities respectively. However, this is not observed in {\fontfamily{lmtt}\selectfont MolHMPN-NoMod} even though we also employ up to 3-hop neighbors for the functional groups as subgraphs. This shows that, rather than increasing the size of the hyperedges, it is more important to select and specify which nodes to put in each hyperedge. For MolSoft, even though it performs well for ClinTox, ESOL and Lipophilicity, it did not perform as well for the rest of the datasets. Hence, from the results, we can verify that it is beneficial to employ prior knowledge 
(in our case, functional groups) to capture substructures that have significant effects on the molecular properties.

\subsection{Hyperedge modification with extended hyperedges}
\label{subsection:extension_ablation}
In Section 1, we made the hypothesis that it is necessary to leverage the prior knowledge \textit{partially} so as to overcome any potentially unsuitable prior knowledge. To test this hypothesis, we use the hyperedge extension and modification scheme. The hyperedges are extended to their $K$-local subgraph as stated in Section 2.1. We refer to {\fontfamily{lmtt}\selectfont MolHMPN} with the $K$-local extension as {\fontfamily{lmtt}\selectfont MolHMPN}-$K$, which also includes the hyperedge modification scheme. We compare {\fontfamily{lmtt}\selectfont MolHMPN-K} with {\fontfamily{lmtt}\selectfont MolHMPN-NoMod}, which uses the functional group hyperedges \textit{without} the modification scheme.

\begin{table}[h]
\centering
\caption{\textbf{Increasing $K$ for hyperedge learning.} Comparison between the different K used. Results in bold are the best-performing results for their respective datasets. ($\uparrow$ means that higher result is better and $\downarrow$ means that lower result is better.)}
\resizebox{\columnwidth}{!}{
\begin{tabular}{l|ccccc|ccc}
\toprule 
Metric & \multicolumn{5}{c|}{AUROC (Classification)} & \multicolumn{3}{c}{RMSE (Regression)} \\
\hline
Dataset
& Tox21 ($\uparrow$) 
& ClinTox ($\uparrow$)
& SIDER ($\uparrow$)
& BBBP ($\uparrow$)
& BACE ($\uparrow$) 
& ESOL ($\downarrow$) 
& FreeSolv ($\downarrow$)
& Lipophilicity ($\downarrow$) \\
\hline 
{\fontfamily{lmtt}\selectfont MolHMPN-NoMod}
& \begin{tabular}[c]{@{}c@{}} 0.839 \\ ($\pm$ 0.0147) \end{tabular}
& \begin{tabular}[c]{@{}c@{}} 0.909 \\ ($\pm$ 0.0394) \end{tabular}
& \begin{tabular}[c]{@{}c@{}} 0.605 \\ ($\pm$ 0.0227) \end{tabular}
& \begin{tabular}[c]{@{}c@{}} 0.928 \\ ($\pm$ 0.0302) \end{tabular}
& \begin{tabular}[c]{@{}c@{}} \textbf{0.892} \\ ($\pm$ 0.0232) \end{tabular}
& \begin{tabular}[c]{@{}c@{}} 0.447 \\ ($\pm$ 0.0559) \end{tabular}
& \begin{tabular}[c]{@{}c@{}} \textbf{0.813} \\ ($\pm$ 0.3952) \end{tabular}
& \begin{tabular}[c]{@{}c@{}} 0.517 \\ ($\pm$ 0.0415) \end{tabular}\\ 
{\fontfamily{lmtt}\selectfont MolHMPN}-0
& \begin{tabular}[c]{@{}c@{}} 0.836 \\ ($\pm$ 0.0146) \end{tabular}
& \begin{tabular}[c]{@{}c@{}} 0.903 \\ ($\pm$ 0.0392) \end{tabular}
& \begin{tabular}[c]{@{}c@{}} 0.596 \\ ($\pm$ 0.0227) \end{tabular}
& \begin{tabular}[c]{@{}c@{}} 0.896 \\ ($\pm$ 0.0299) \end{tabular}
& \begin{tabular}[c]{@{}c@{}} 0.856 \\ ($\pm$ 0.0232) \end{tabular}
& \begin{tabular}[c]{@{}c@{}} 0.454 \\ ($\pm$ 0.0339) \end{tabular}
& \begin{tabular}[c]{@{}c@{}} 1.020 \\ ($\pm$ 0.3606) \end{tabular}
& \begin{tabular}[c]{@{}c@{}} 0.535 \\ ($\pm$ 0.0391) \end{tabular}\\ 
{\fontfamily{lmtt}\selectfont MolHMPN}-1
& \begin{tabular}[c]{@{}c@{}} \textbf{0.840} \\ ($\pm$ 0.0157) \end{tabular}
& \begin{tabular}[c]{@{}c@{}} 0.906 \\ ($\pm$ 0.0362) \end{tabular}
& \begin{tabular}[c]{@{}c@{}} 0.605 \\ ($\pm$ 0.0124) \end{tabular}
& \begin{tabular}[c]{@{}c@{}} \textbf{0.940} \\ ($\pm$ 0.0227) \end{tabular}
& \begin{tabular}[c]{@{}c@{}} 0.884 \\ ($\pm$ 0.0261) \end{tabular}
& \begin{tabular}[c]{@{}c@{}} 0.404 \\ ($\pm$ 0.0672) \end{tabular}
& \begin{tabular}[c]{@{}c@{}} 1.090 \\ ($\pm$ 0.5495) \end{tabular}
& \begin{tabular}[c]{@{}c@{}} \textbf{0.514} \\ ($\pm$ 0.1037) \end{tabular}\\
{\fontfamily{lmtt}\selectfont MolHMPN}-2
& \begin{tabular}[c]{@{}c@{}} 0.839 \\ ($\pm$ 0.0163) \end{tabular}
& \begin{tabular}[c]{@{}c@{}} \textbf{0.919} \\ ($\pm$ 0.0485) \end{tabular}
& \begin{tabular}[c]{@{}c@{}} \textbf{0.617} \\ ($\pm$ 0.0160) \end{tabular}
& \begin{tabular}[c]{@{}c@{}} 0.905 \\ ($\pm$ 0.0298) \end{tabular}
& \begin{tabular}[c]{@{}c@{}} 0.886 \\ ($\pm$ 0.0175) \end{tabular}
& \begin{tabular}[c]{@{}c@{}} \textbf{0.390} \\ ($\pm$ 0.0492) \end{tabular}
& \begin{tabular}[c]{@{}c@{}} 1.471 \\ ($\pm$ 0.4929) \end{tabular}
& \begin{tabular}[c]{@{}c@{}} 0.619 \\ ($\pm$ 0.1104) \end{tabular}\\ 

\bottomrule
\end{tabular}
}
\label{tab:increasek}
\end{table}
\begin{figure}
    \centering
    \begin{subfigure}[b]{0.24\textwidth}
    \includegraphics[width={1.2\linewidth}]{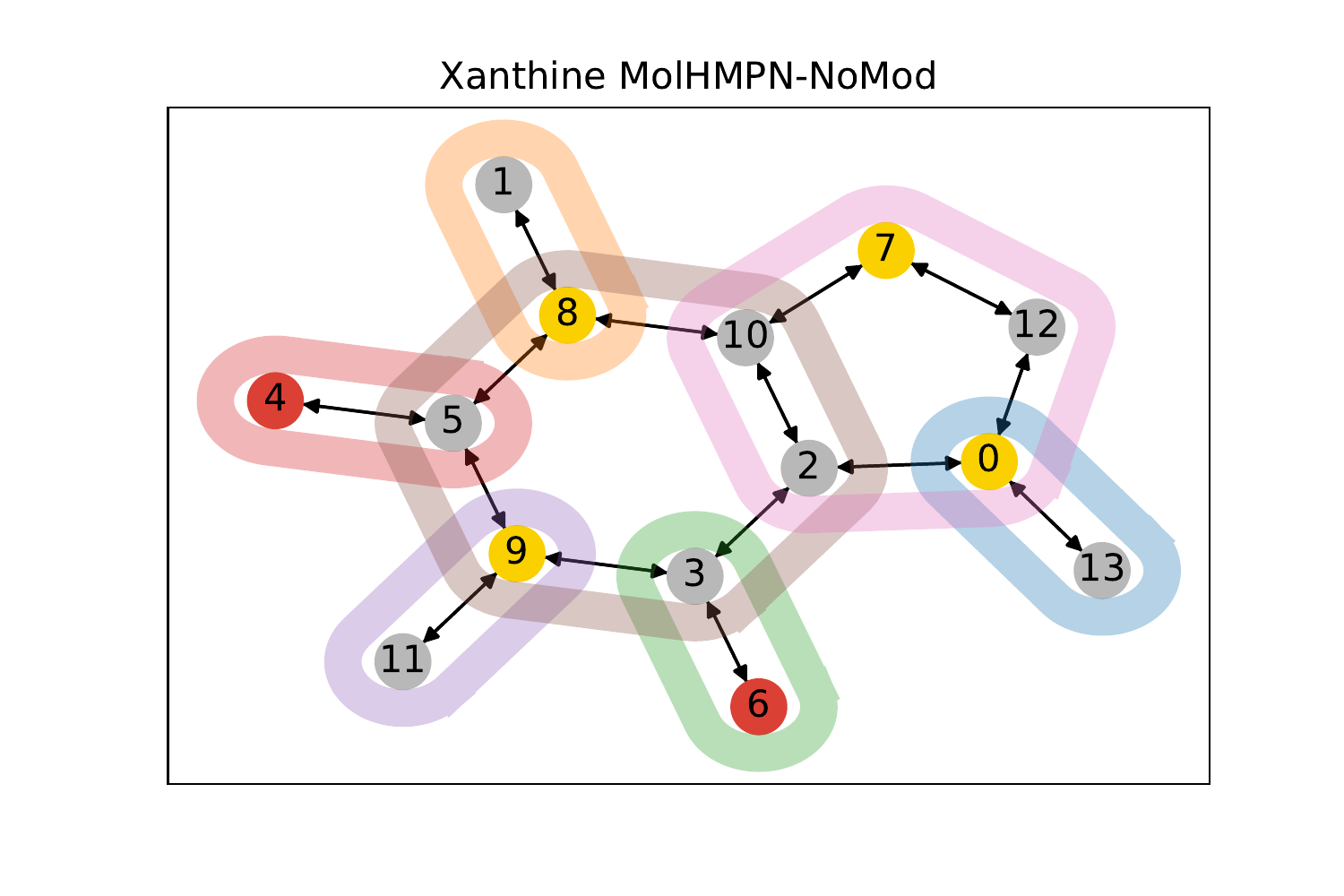}
    \end{subfigure}
    \begin{subfigure}[b]{0.24\textwidth}
    \includegraphics[width={1.2\linewidth}]{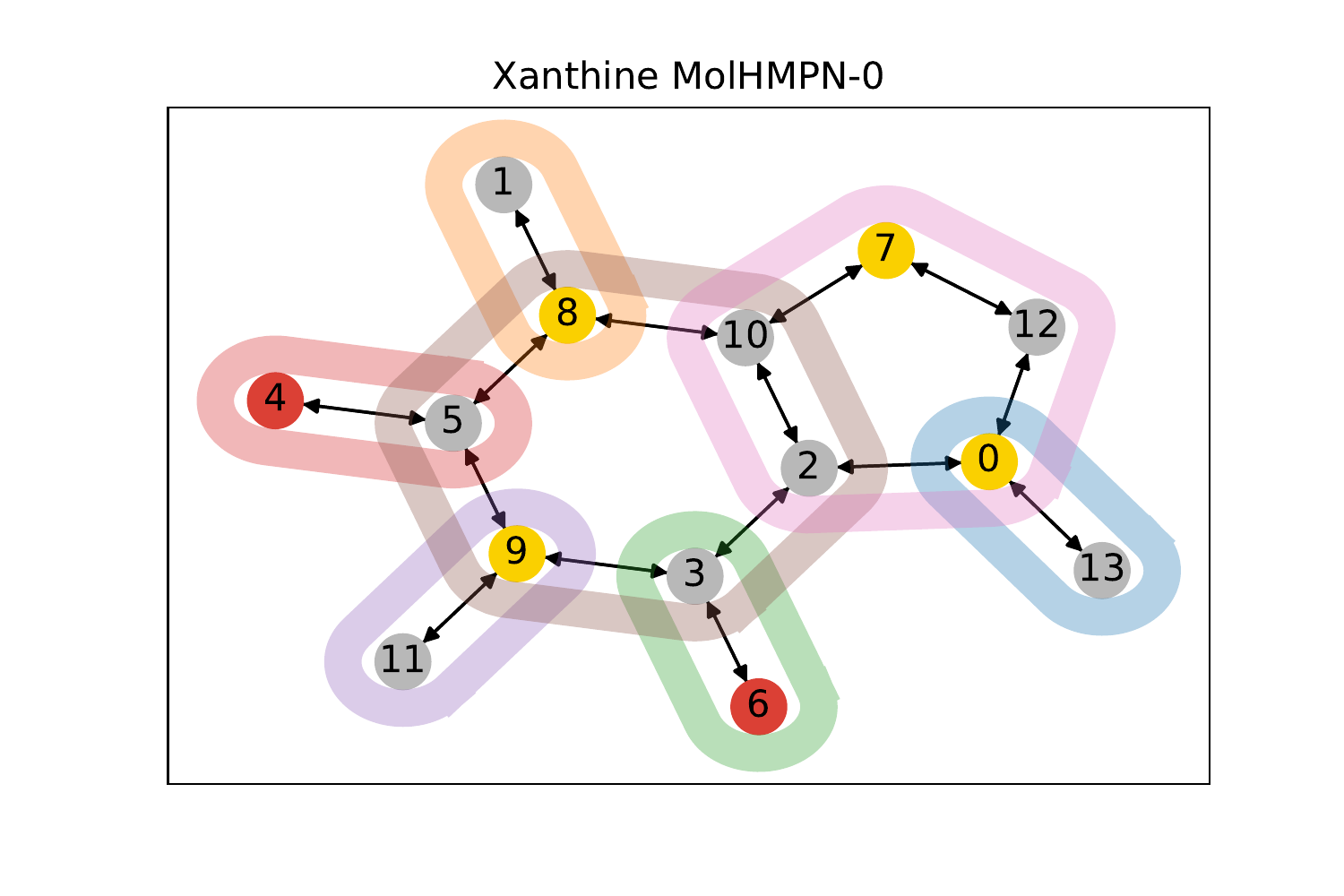}
    \end{subfigure}
    \begin{subfigure}[b]{0.24\textwidth}
    \includegraphics[width={1.2\linewidth}]{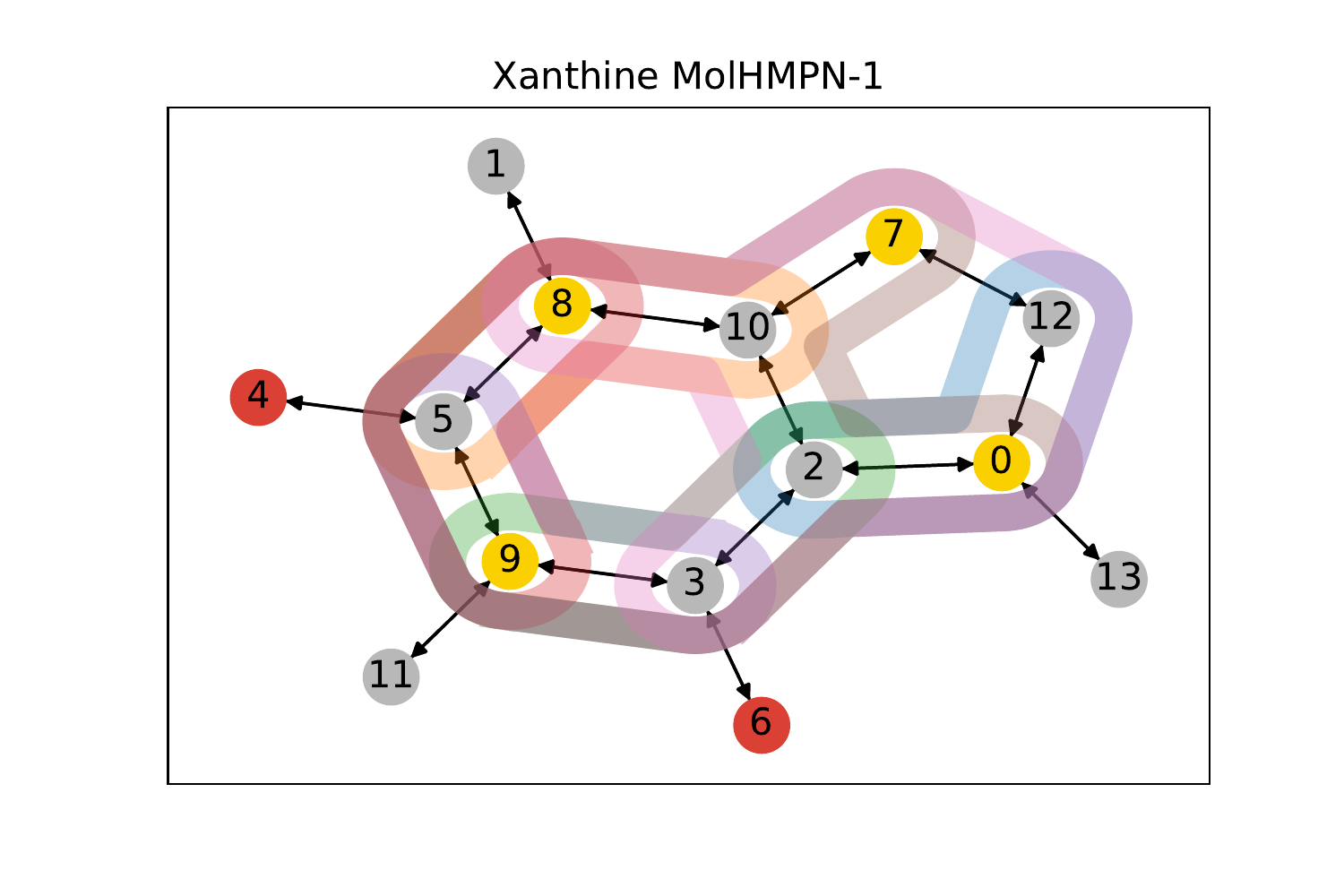}
    \end{subfigure}
    \begin{subfigure}[b]{0.24\textwidth}
    \includegraphics[width={1.2\linewidth}]{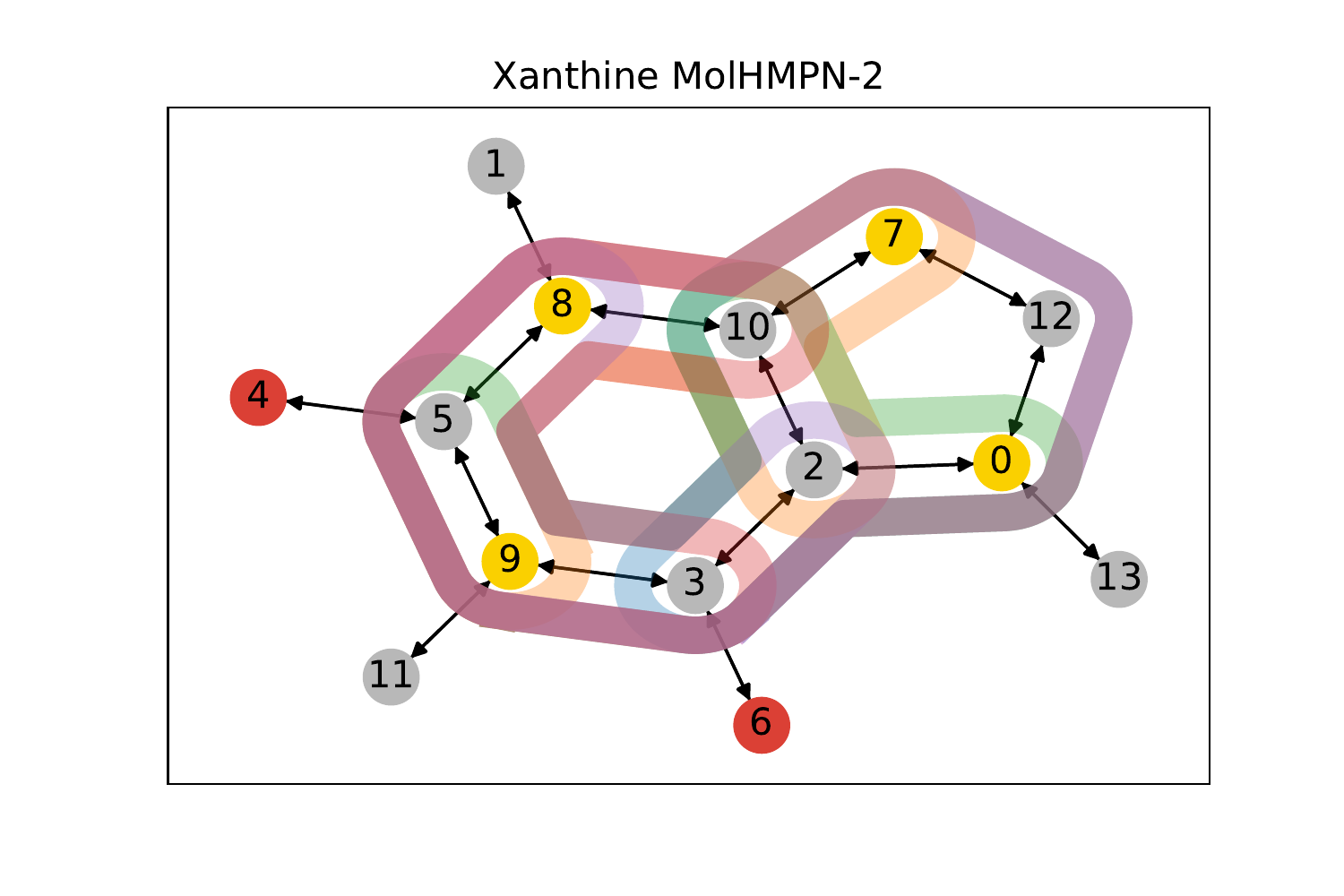}
    \end{subfigure}
    \begin{subfigure}[b]{0.24\textwidth}
    \includegraphics[width={1.2\linewidth}]{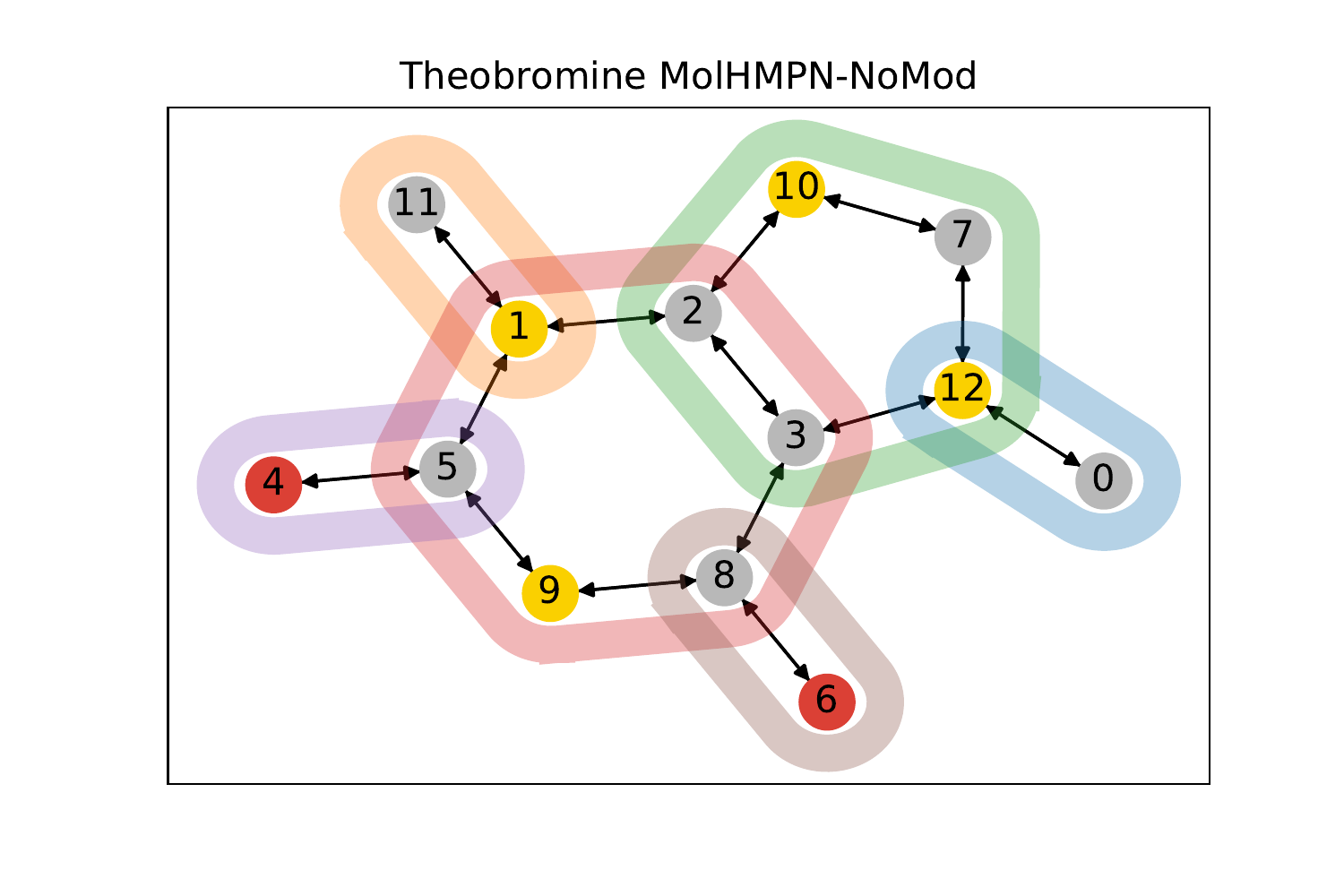}
    \end{subfigure}
    \begin{subfigure}[b]{0.24\textwidth}
    \includegraphics[width={1.2\linewidth}]{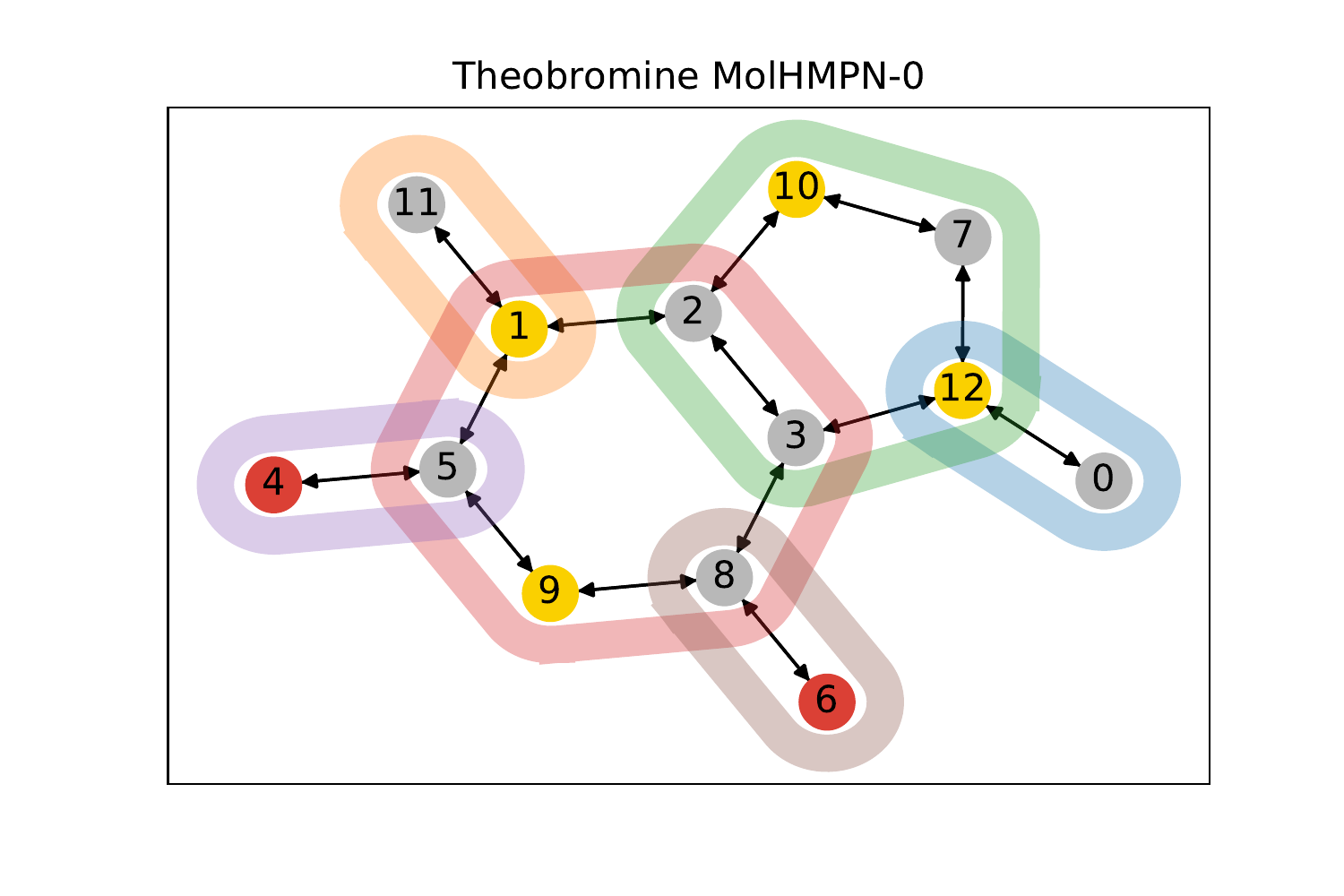}
    \end{subfigure}
    \begin{subfigure}[b]{0.24\textwidth}
    \includegraphics[width={1.2\linewidth}]{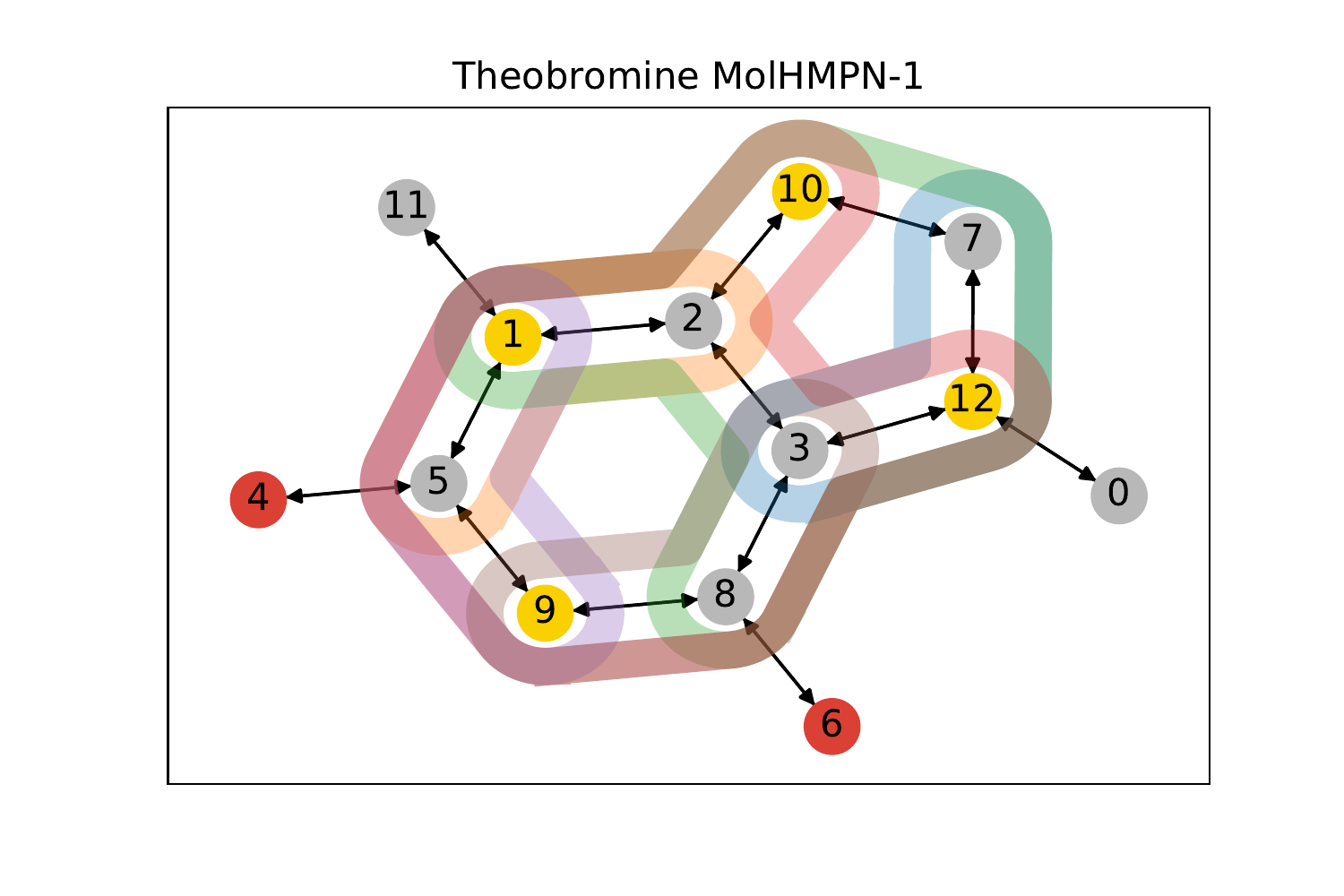}
    \end{subfigure}
    \begin{subfigure}[b]{0.24\textwidth}
    \includegraphics[width={1.2\linewidth}]{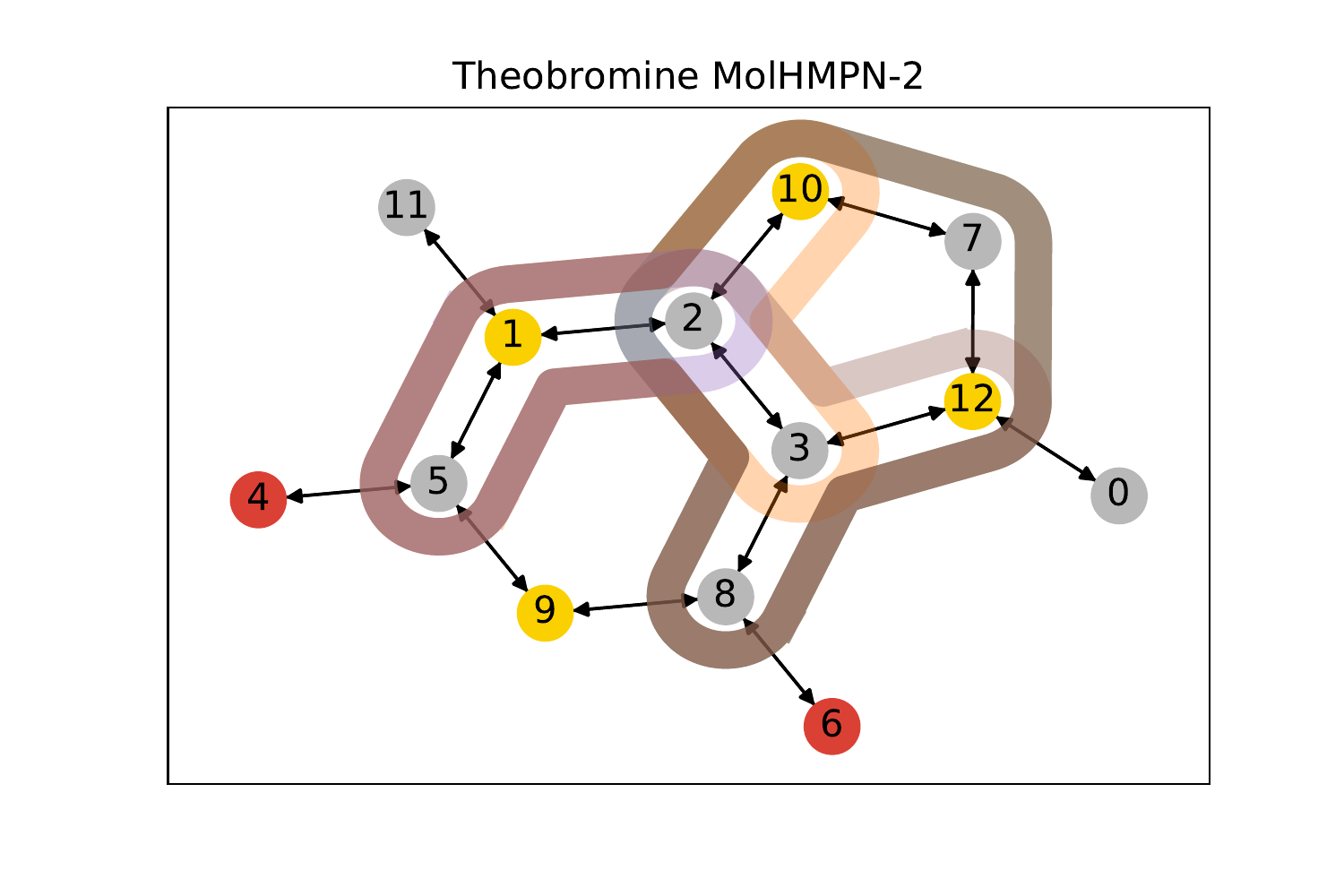}
    \end{subfigure}
    \begin{subfigure}[b]{0.24\textwidth}
    \includegraphics[width={1.2\linewidth}]{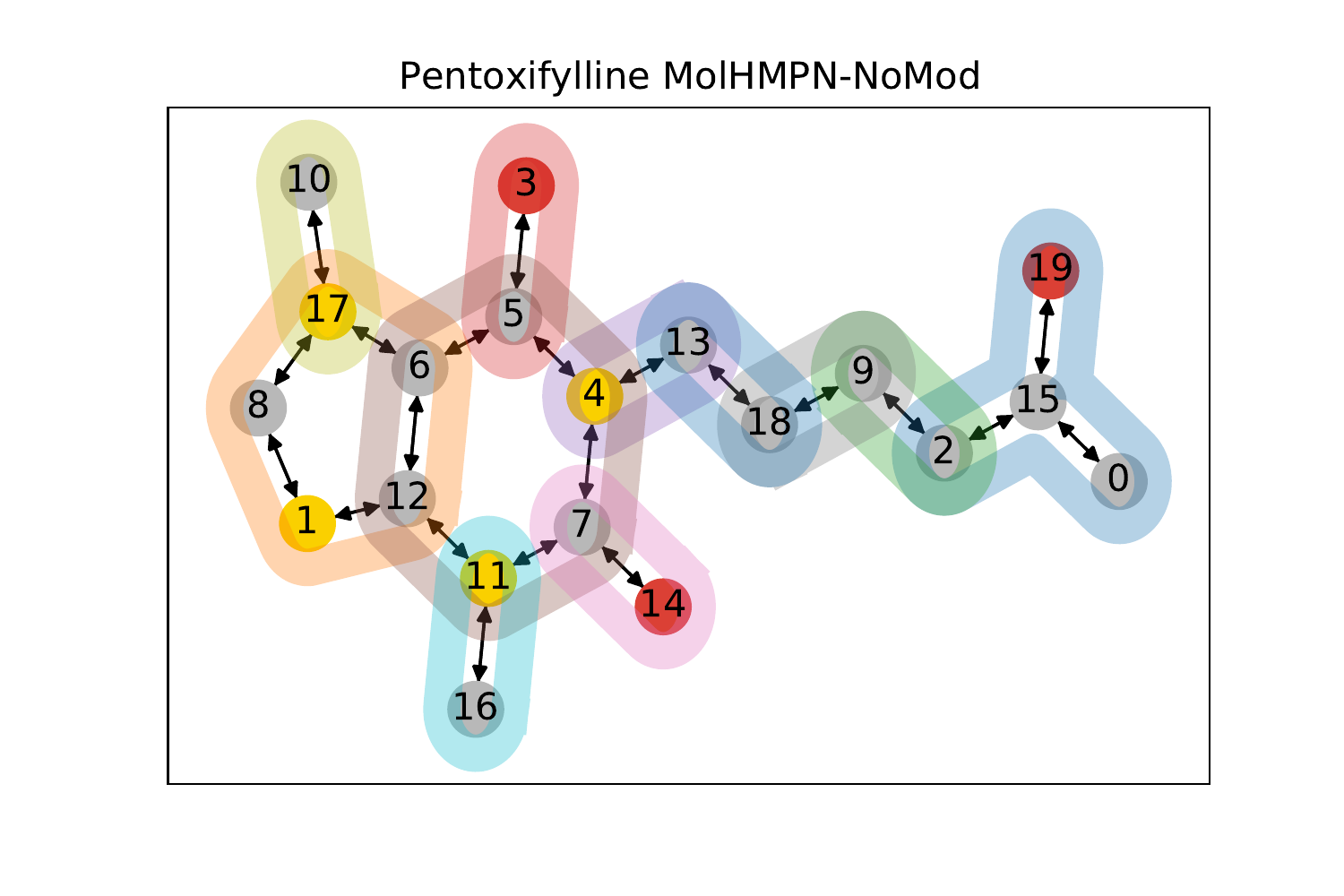}
    \end{subfigure}
    \begin{subfigure}[b]{0.24\textwidth}
    \includegraphics[width={1.2\linewidth}]{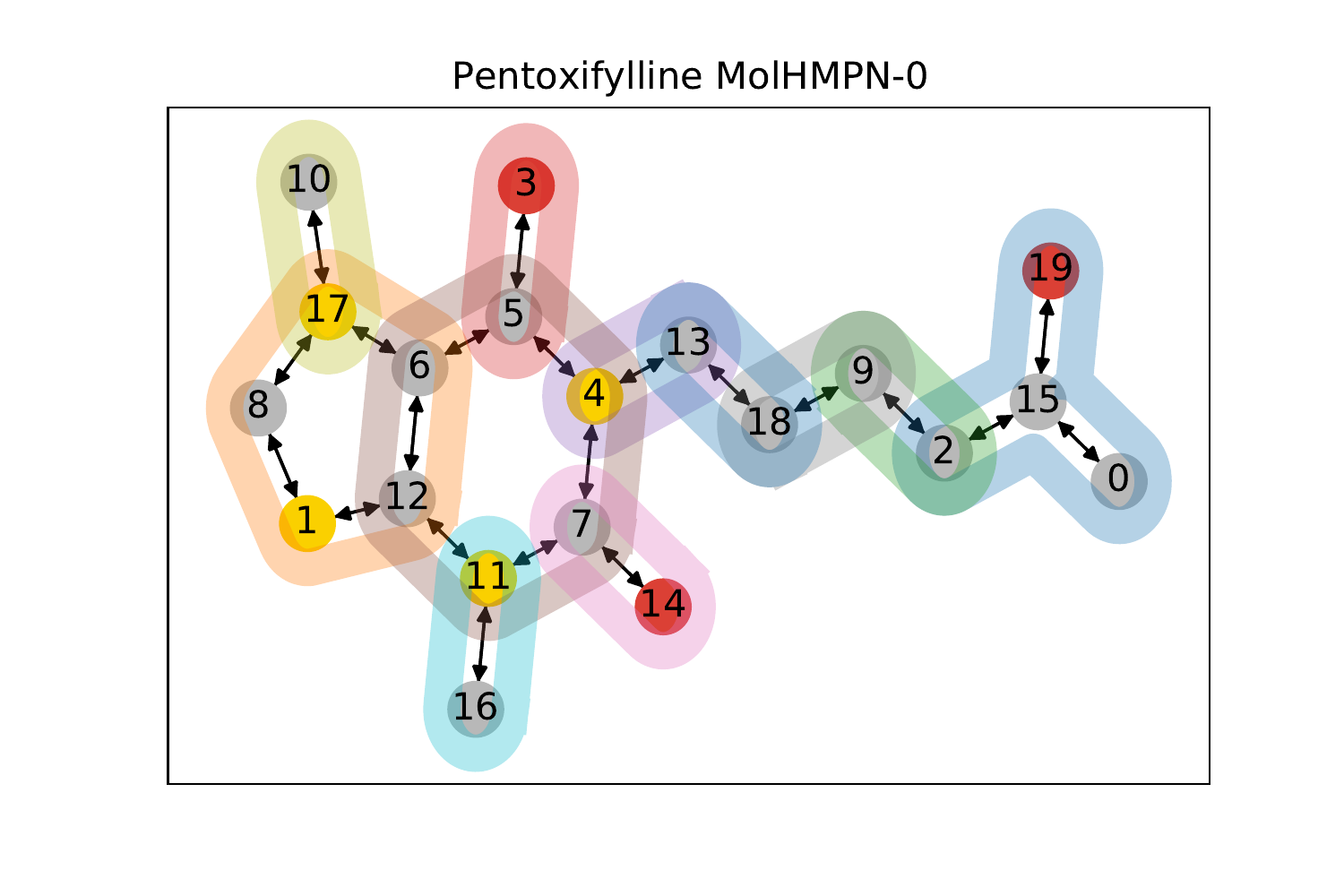}
    \end{subfigure}
    \begin{subfigure}[b]{0.24\textwidth}
    \includegraphics[width={1.2\linewidth}]{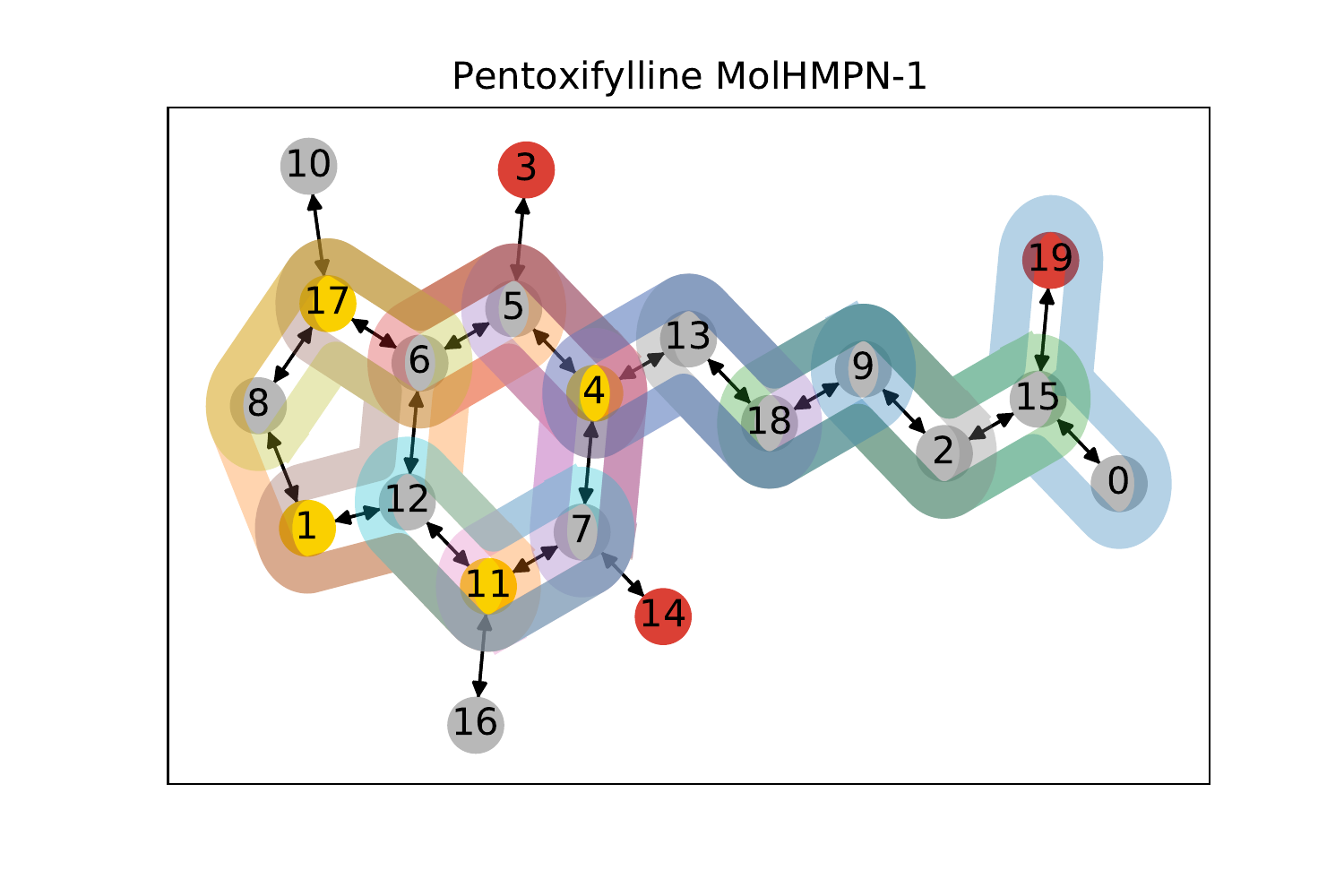}
    \end{subfigure}
    \begin{subfigure}[b]{0.24\textwidth}
    \includegraphics[width={1.2\linewidth}]{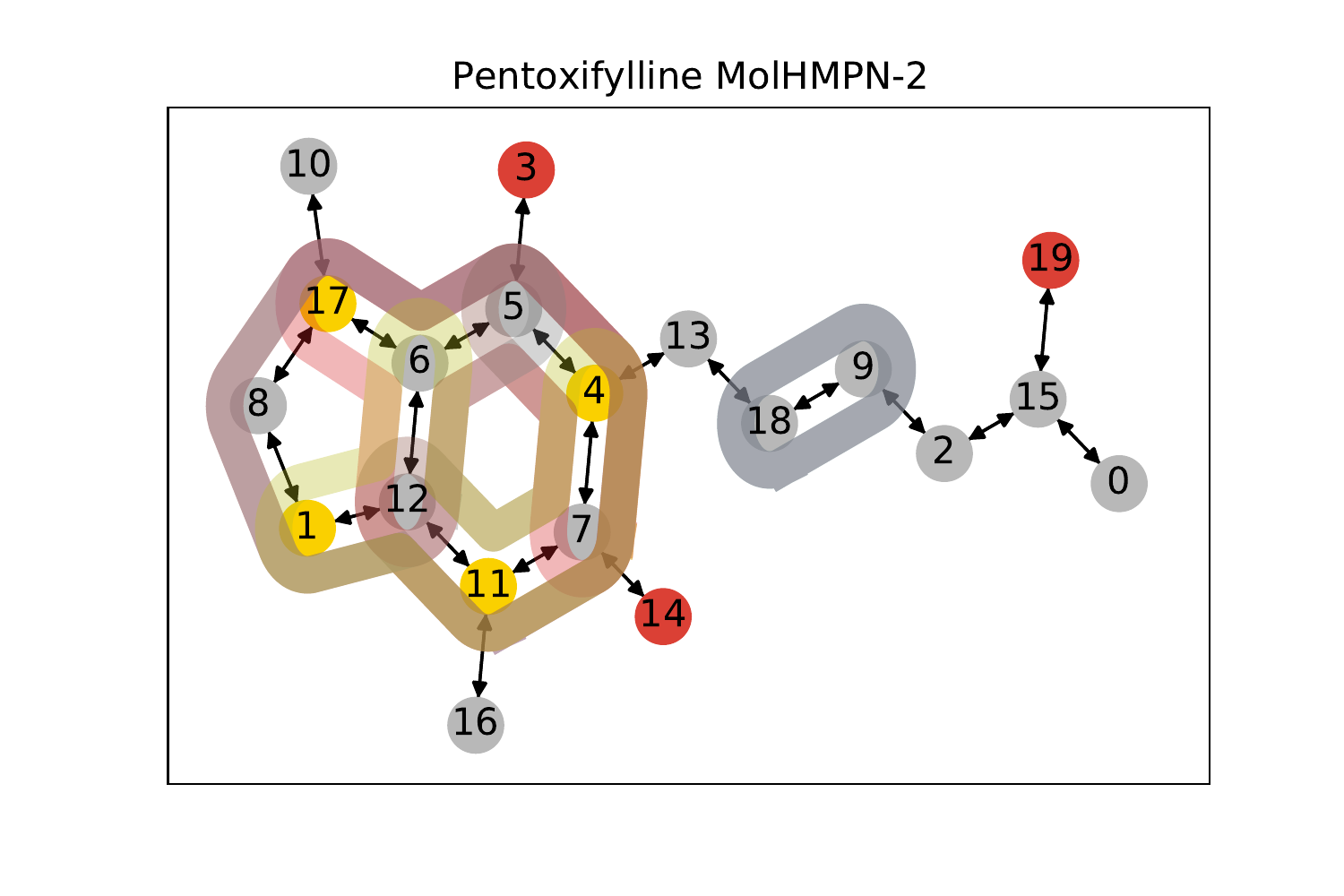}
    \end{subfigure}
    \caption{\textbf{Hyperedge visualization}. Comparing between the hyperedges in {\fontfamily{lmtt}\selectfont MolHMPN-NoMod} and {\fontfamily{lmtt}\selectfont MolHMPN}-$K$.} 
    \label{fig:molext}
\end{figure}

Table \ref{tab:increasek} shows the results of the effects of increasing $K$. From Table \ref{tab:increasek}, we can see that the extension and modification scheme has improved the performance of {\fontfamily{lmtt}\selectfont MolHMPN-NoMod} generally when $K \geq 1$. Among the datasets, FreeSolv performs significantly better in {\fontfamily{lmtt}\selectfont MolHMPN-NoMod} than in {\fontfamily{lmtt}\selectfont MolHMPN}-K. This is supported by the fact that FreeSolv contains fragment-like compounds where the typical size of the molecules is substantially smaller than typical small-molecule drugs \cite{mobley2014freesolv}. Large performance degradation is observed for BBBP and Lipophilicity from {\fontfamily{lmtt}\selectfont MolHMPN}-1 to {\fontfamily{lmtt}\selectfont MolHMPN}-2. This may be because the extended hyperedges have deviated too far away from the original functional group design for the two datasets, and have most parts overlapped with another hyperedge, making the hyperedges indistinguishable. To have a qualitative analysis of how $K$ affects the performances, we provide a visualization of how the hyperedge changes when $K$ changes.

We revisit xanthine, theobromine and pentoxifylline in \Figref{fig:molcompare} to visualize the changes in the hyperedges when $K$ changes. Figure \ref{fig:molext} shows an example of how the hyperedges change when $K$ changes. From \Figref{fig:molext}, no changes from the hyperedges were observed in {\fontfamily{lmtt}\selectfont MolHMPN}-0 and the performance drop from {\fontfamily{lmtt}\selectfont MolHMPN-NoMod} may be due to the difference between the {\fontfamily{lmtt}\selectfont MolHMPN} models.
When $K=1,2$, we can see that most parts of the purine ring (i.e. the two cyclic groups) has been covered in all three molecules, even after the hyperedge modification. The purine ring is identified as one of the most frequently-occuring substructures in drugs in the Comprehensive Medicinal Chemistry (CMC) database \cite{guy:1996}. The substructure in pentoxifylline that is not observed in xanthine and theobromine (i.e. ketone and carbon chain) is also covered when $K=1$. This shows that the model has captured the substructures that affects the properties of the molecules successfully.  

\begin{figure}
    \centering
    \begin{subfigure}[t]{0.49\textwidth}
    \includegraphics[width={1.2\linewidth}]{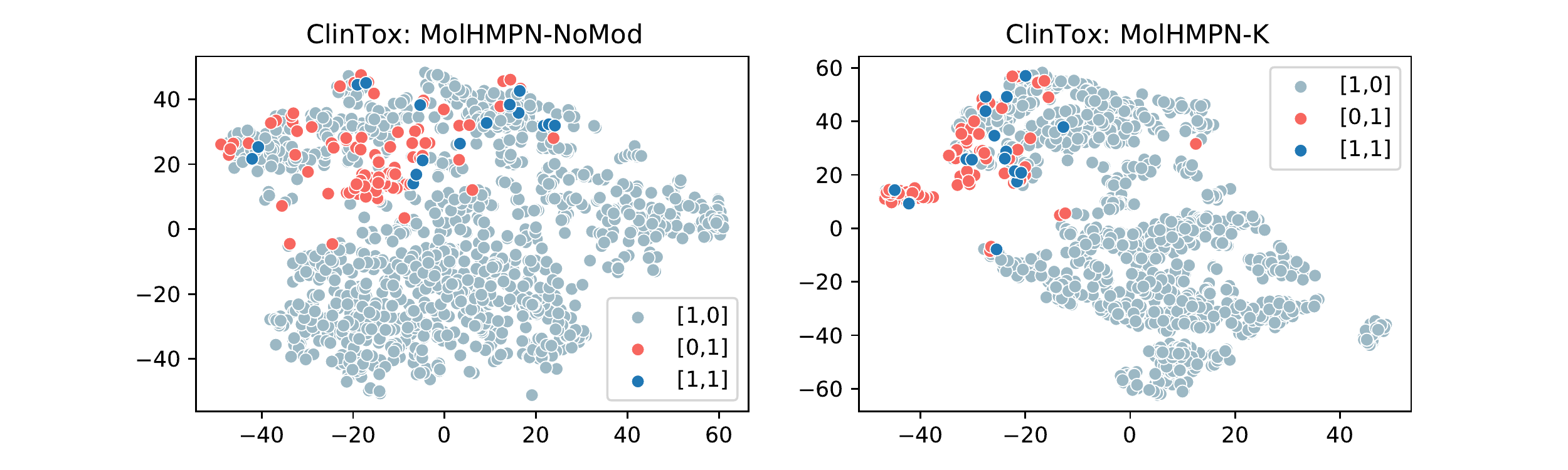}
    \end{subfigure}
    \begin{subfigure}[t]{0.49\textwidth}
    \includegraphics[width={1.2\linewidth}]{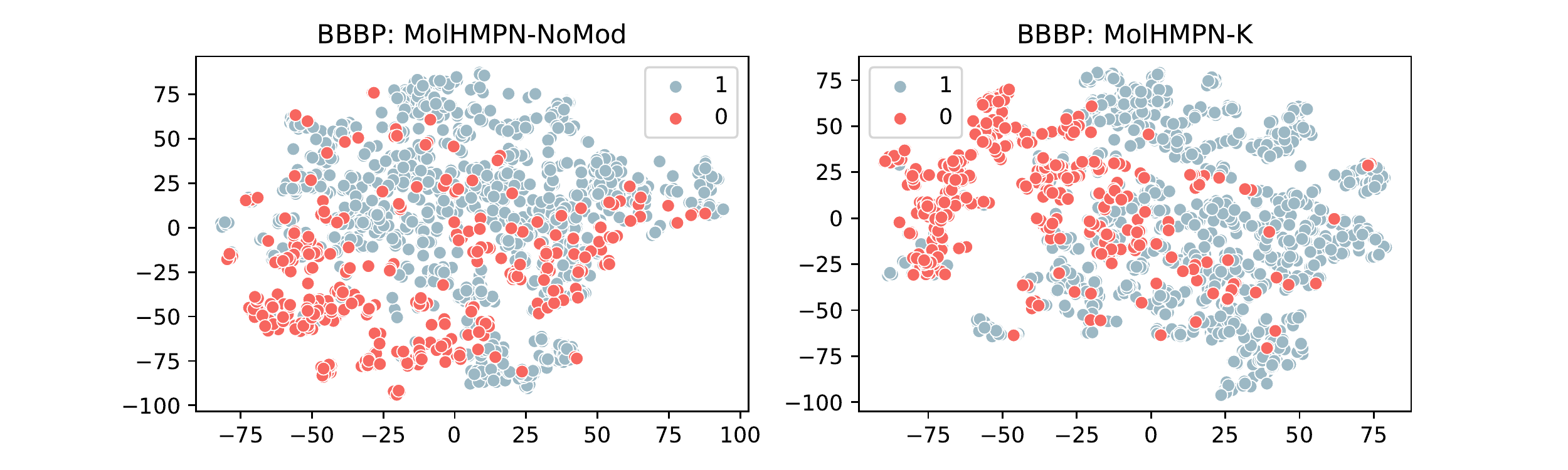}
    \end{subfigure}
    \begin{subfigure}[b]{0.49\textwidth}
    \includegraphics[width={1.2\linewidth}]{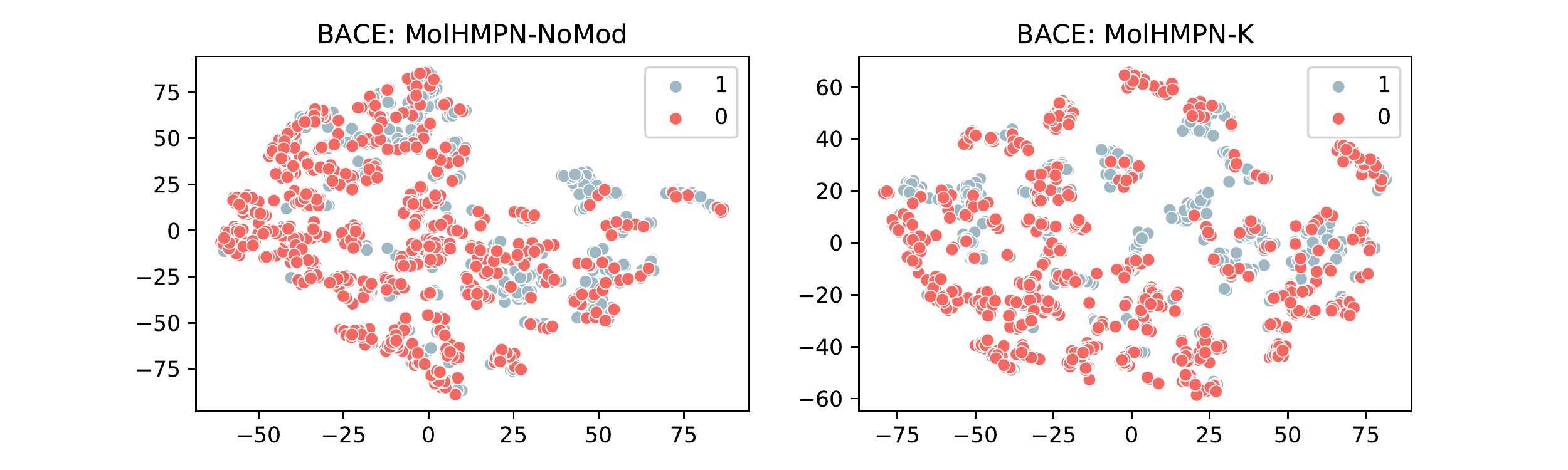}
    \end{subfigure}
    \begin{subfigure}[b]{0.49\textwidth}
    \includegraphics[width={1.2\linewidth}]{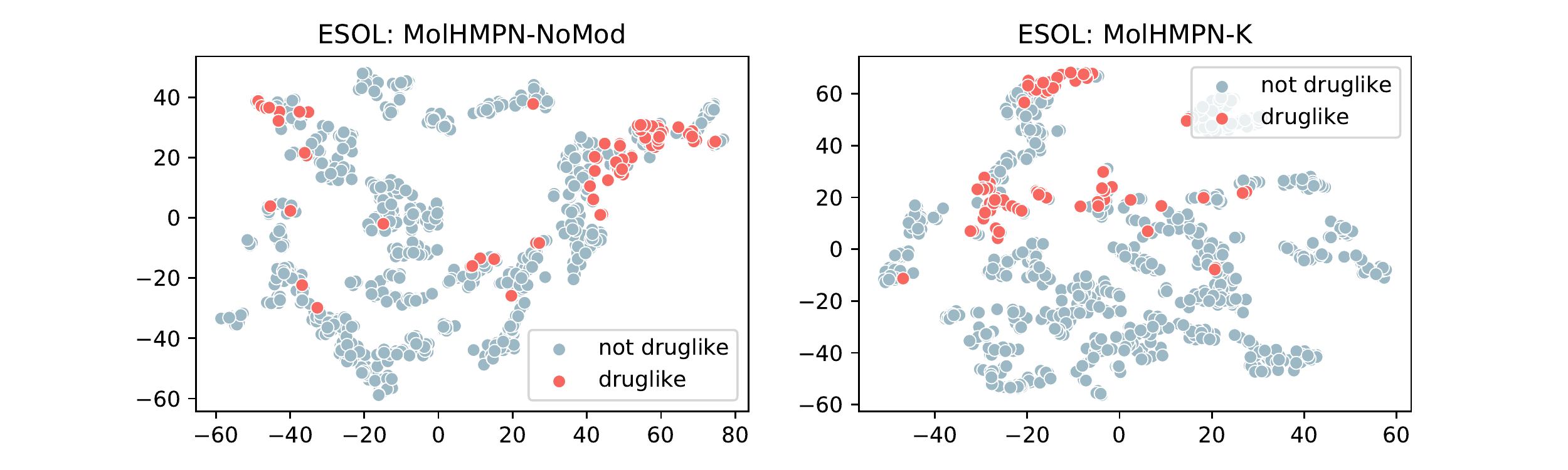}
    \end{subfigure}
    \begin{subfigure}[b]{0.49\textwidth}
    \includegraphics[width={1.2\linewidth}]{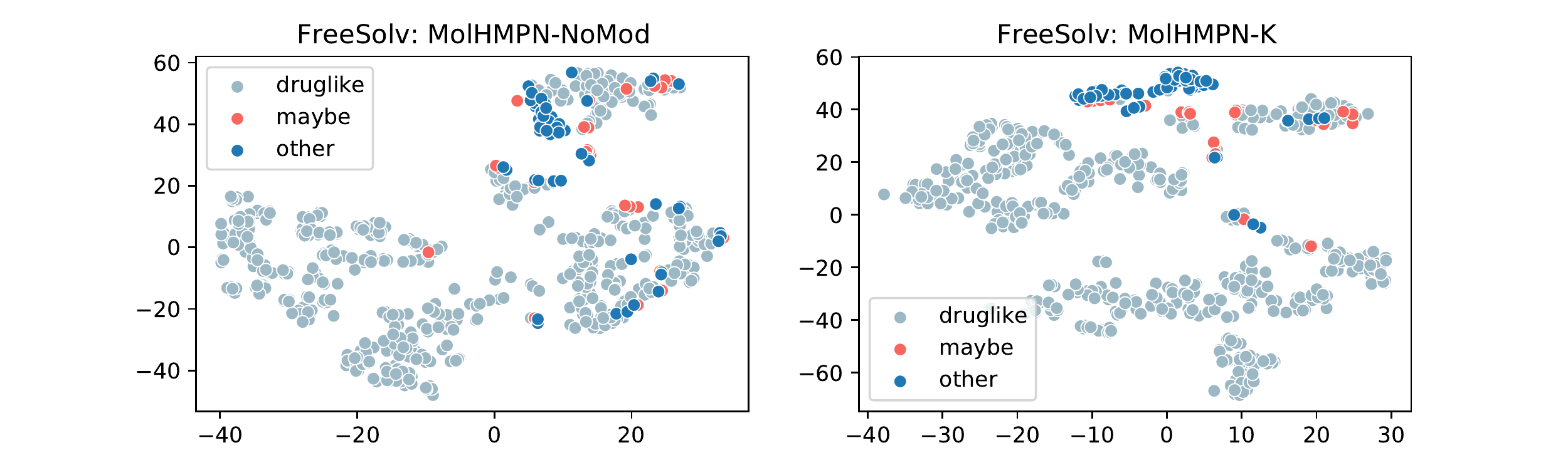}
    \end{subfigure}
    \begin{subfigure}[b]{0.49\textwidth}
    \includegraphics[width={1.2\linewidth}]{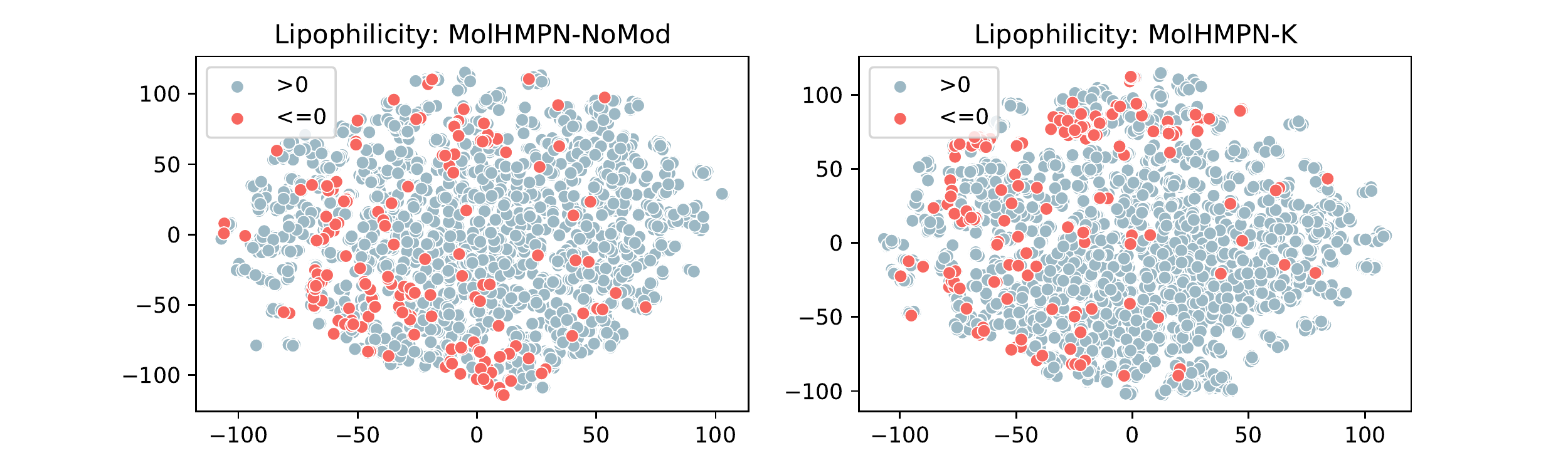}
    \end{subfigure}
    \caption{\textbf{t-SNE plots for {\fontfamily{lmtt}\selectfont MolHMPN-NoMod} and {\fontfamily{lmtt}\selectfont MolHMPN}-$K$.} Analyzing the contributions of the hyperedge extension and modification scheme.}
    \label{fig:ext_tsne}
\end{figure}

We also provide the t-SNE plots for {\fontfamily{lmtt}\selectfont MolHMPN-NoMod} and {\fontfamily{lmtt}\selectfont MolHMPN}-$K$. \Figref{fig:ext_tsne} shows the t-SNE plots for {\fontfamily{lmtt}\selectfont MolHMPN-NoMod} and {\fontfamily{lmtt}\selectfont MolHMPN}-$K$. From the BACE and Lipophilicity plots, we again do not see a clear separation between the data types. However, from the ClinTox, BBBP, ESOL and FreeSolv plots, we can see that the hyperedge extension and modification have either improved or maintained clear separability between different data types generally. 

Overall, from these results, we can verify that leveraging the prior knowledge \textit{partially} is beneficial for the molecular properties prediction.

\section{Conclusion}
We propose a molecular hyper-message passing network ({\fontfamily{lmtt}\selectfont MolHMPN}) to integrate pair-wise and higher-order connectivities in molecules, and using domain knowledge-guided learnt substructures for molecular properties prediction tasks. 
We construct the hypergraph representation of molecules using functional groups, embed the graphs and constructed hypergraphs using the  {\fontfamily{lmtt}\selectfont HyperMP} layer, modify the membership of the hyperedges using the computed embeddings, and predict the molecular properties using the embeddings of the graphs and modified hypergraphs. We evaluate the performance of our model with several baseline methods, and show that our model is able to achieve outstanding results with only one {\fontfamily{lmtt}\selectfont HyperMP} layer. We also analyze our design choices by comparing the results of AtomGC and FuncGC with {\fontfamily{lmtt}\selectfont MolHMPN-NoMod}, analyze the 
performance of {\fontfamily{lmtt}\selectfont MolHMPN} with increased $K$ in hyperedge learning, and compare the performances using different types of substructures. We have verified that the incorporation of prior knowledge improves the performances and that the domain knowledge-guided hyperedge modification plays a crucial role when modeling higher-order connectivities robustly.

\newpage
\bibliography{iclr2022_conference}
\bibliographystyle{iclr2022_conference}


\newpage
\appendix
\section{Appendix}
\setcounter{figure}{0}
\setcounter{table}{0}
\counterwithin{figure}{section}
\counterwithin{table}{section}
\subsection{Hypergraph construction}
\label{appen:hyperconstr}
In this section, we provide the list of functional groups that have been utilized in our current study based on their central atoms. We highlight the central atoms and their respective first- and second-hop neighbors with circles of different colors.

\begin{table}[ht]
\caption{\textbf{Functional groups with nitrogen as the central atom.} The red circles represent the central atoms, and the blue and green circles represent the 1-hop and 2-hop neighbors from the central atom respectively.}
\begin{tabular}{ccc|ccc}
\toprule
Functional group & Structure & Hyperedge & Functional group & Structure & Hyperedge\\
\hline 
Amine & 
\includegraphics[width={.10\linewidth}, trim=0 0 0 -5]{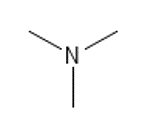} &  \includegraphics[width={.10\linewidth}]{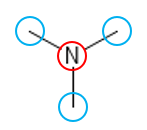} & 
Nitro & 
\includegraphics[width={.09\linewidth}]{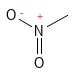} &  \includegraphics[width={.09\linewidth}]{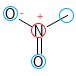}\\
\hline 
Nitrate & 
\includegraphics[width={.10\linewidth}, trim=0 0 0 -5]{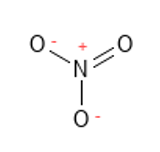} &  \includegraphics[width={.10\linewidth}]{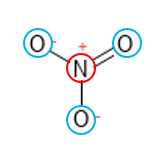} & 
C nitroso & 
\includegraphics[width={.09\linewidth}]{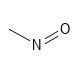} &  \includegraphics[width={.09\linewidth}]{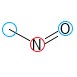}\\
\hline 
N nitroso & 
\includegraphics[width={.11\linewidth}, trim=0 0 0 -5]{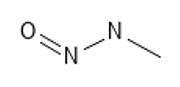} &  \includegraphics[width={.11\linewidth}]{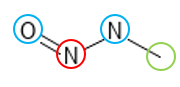} & 
Azo & 
\includegraphics[width={.09\linewidth}]{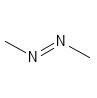} &  \includegraphics[width={.09\linewidth}]{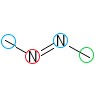}\\
\hline 
Hydrazine &
\includegraphics[width={.10\linewidth}, trim=0 0 0 -5]{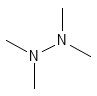} &  \includegraphics[width={.10\linewidth}]{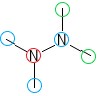} & 
Hydroxylamine & 
\includegraphics[width={.09\linewidth}]{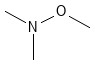} &  \includegraphics[width={.09\linewidth}]{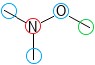}\\
\hline
Nitrile &
\includegraphics[width={.10\linewidth}, trim=0 0 0 -5]{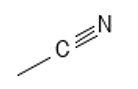} &  \includegraphics[width={.10\linewidth}]{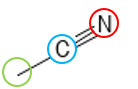}\\
\bottomrule
\end{tabular}
\end{table}

\begin{table}[ht]
\caption{\textbf{Functional groups with carbon as the central atom.} The red circles represent the central atoms, and the blue and green circles represent the 1-hop and 2-hop neighbors from the central atom respectively.}
\begin{tabular}{ccc|ccc}
\toprule
Functional group & Structure & Hyperedge & Functional group & Structure & Hyperedge\\
\hline 
Alkene & 
\includegraphics[width={.10\linewidth}, trim=0 0 0 -5]{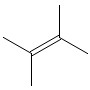} &  \includegraphics[width={.10\linewidth}]{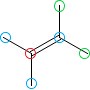} & 
Alkyne & 
\includegraphics[width={.11\linewidth}]{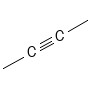} &  \includegraphics[width={.11\linewidth}]{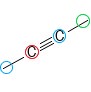}\\
\hline
Aldehyde & 
\includegraphics[width={.085\linewidth}, trim=0 0 0 -5]{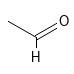} &  \includegraphics[width={0.085\linewidth}]{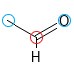} & 
Ketene & 
\includegraphics[width={.11\linewidth}]{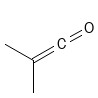} &  \includegraphics[width={.11\linewidth}]{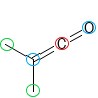}\\
\hline
Isocynate & 
\includegraphics[width={.12\linewidth}, trim=0 0 0 -5]{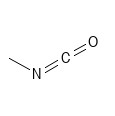} &  \includegraphics[width={.12\linewidth}]{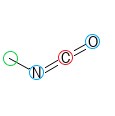} & 
Carboxyl & 
\includegraphics[width={.10\linewidth}]{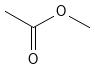} &  \includegraphics[width={.1\linewidth}]{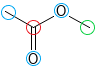}\\
\hline
Carbamate & 
\includegraphics[width={.1\linewidth}, trim=0 0 0 -5]{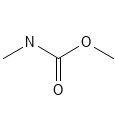} &  \includegraphics[width={.1\linewidth}]{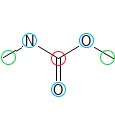} & 
Carbamide & 
\includegraphics[width={.105\linewidth}]{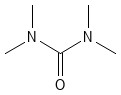} &  \includegraphics[width={.105\linewidth}]{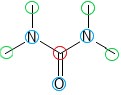}\\
\hline
Amide & 
\includegraphics[width={.09\linewidth}, trim=0 0 0 -5]{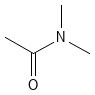} &  \includegraphics[width={.09\linewidth}]{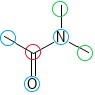} & 
Ketone & 
\includegraphics[width={.09\linewidth}]{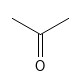} &  \includegraphics[width={.09\linewidth}]{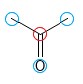}\\
\hline
Isothiocynate & 
\includegraphics[width={.1\linewidth}, trim=0 0 0 -5]{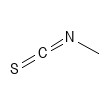} &  \includegraphics[width={.1\linewidth}]{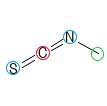} & 
Thione & 
\includegraphics[width={.1\linewidth}]{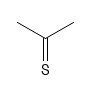} &  \includegraphics[width={.1\linewidth}]{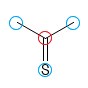}\\
\hline
Thioamide & 
\includegraphics[width={.1\linewidth}, trim=0 0 0 -5]{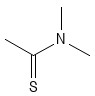} &  \includegraphics[width={.1\linewidth}]{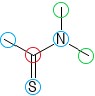} & 
Thiourea & 
\includegraphics[width={.1\linewidth}]{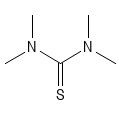} &  \includegraphics[width={.1\linewidth}]{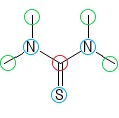}\\
\hline
Carbodiimide & 
\includegraphics[width={.11\linewidth}, trim=0 0 0 -5]{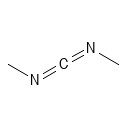} &  \includegraphics[width={.11\linewidth}]{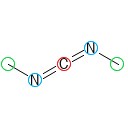} & 
Carboximidamide & 
\includegraphics[width={.1\linewidth}]{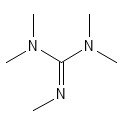} &  \includegraphics[width={.1\linewidth}]{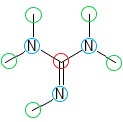}\\
\hline
Imine & 
\includegraphics[width={.095\linewidth}, trim=0 0 0 -5]{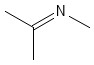} &  \includegraphics[width={.095\linewidth}]{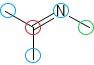} & 
Hydrazone & 
\includegraphics[width={.11\linewidth}]{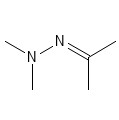} &  \includegraphics[width={.11\linewidth}]{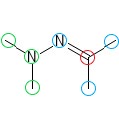}\\
\hline
Oxime & 
\includegraphics[width={.11\linewidth}, trim=0 0 0 -5]{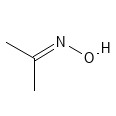} &  \includegraphics[width={.11\linewidth}]{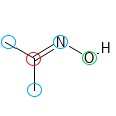} & 
Alcohol & 
\includegraphics[width={.1\linewidth}]{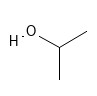} &  \includegraphics[width={.1\linewidth}]{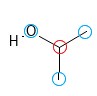}\\
\hline
Thiol & 
\includegraphics[width={.11\linewidth}, trim=0 0 0 -5]{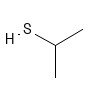} &  \includegraphics[width={.11\linewidth}]{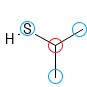} & 
Allene & 
\includegraphics[width={.1\linewidth}]{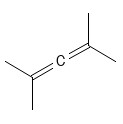} &  \includegraphics[width={.1\linewidth}]{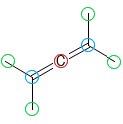}\\
\bottomrule
\end{tabular}
\end{table}

\begin{table}[ht]
\caption{\textbf{Functional groups with oxygen as the central atom.} The red circles represent the central atoms, and the blue and green circles represent the 1-hop and 2-hop neighbors from the central atom respectively.}
\begin{tabular}{ccc|ccc}
\toprule
Functional group & Structure & Hyperedge & Functional group & Structure & Hyperedge\\
\hline 
Ether & 
\includegraphics[width={.10\linewidth}, trim=0 0 0 -5]{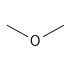} &  \includegraphics[width={.10\linewidth}]{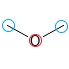} & 
Peroxide & 
\includegraphics[width={.11\linewidth}]{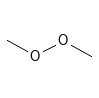} &  \includegraphics[width={.11\linewidth}]{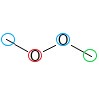}\\
\bottomrule
\end{tabular}
\end{table}

\begin{table}[ht]
\caption{\textbf{Functional groups with phosphorus as the central atom.} The red circles represent the central atoms, and the blue and green circles represent the 1-hop and 2-hop neighbors from the central atom respectively.}
\begin{tabular}{ccc|ccc}
\toprule
Functional group & Structure & Hyperedge & Functional group & Structure & Hyperedge\\
\hline 
Phosphanyl & 
\includegraphics[width={.10\linewidth}, trim=0 0 0 -5]{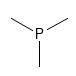} &  \includegraphics[width={.10\linewidth}]{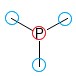} & 
Phosphine oxide & 
\includegraphics[width={.11\linewidth}]{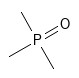} &  \includegraphics[width={.11\linewidth}]{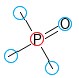}\\
\hline 
Phosphite ester & 
\includegraphics[width={.13\linewidth}, trim=0 0 0 -5]{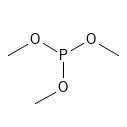} &  \includegraphics[width={.13\linewidth}]{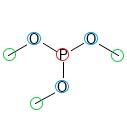} & 
Phosphodiester & 
\includegraphics[width={.11\linewidth}]{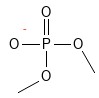} &  \includegraphics[width={.11\linewidth}]{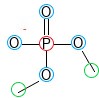}\\
\bottomrule
\end{tabular}
\end{table}

\begin{table}[ht]
\caption{\textbf{Functional groups with sulfur as the central atom.} The red circles represent the central atoms, and the blue and green circles represent the 1-hop and 2-hop neighbors from the central atom respectively.}
\begin{tabular}{ccc|ccc}
\toprule
Functional group & Structure & Hyperedge & Functional group & Structure & Hyperedge\\
\hline 
Disulfide & 
\includegraphics[width={.11\linewidth}, trim=0 0 0 -5]{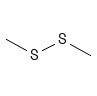} &  \includegraphics[width={.11\linewidth}]{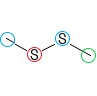} & 
Sulfoxide & 
\includegraphics[width={.11\linewidth}]{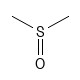} &  \includegraphics[width={.11\linewidth}]{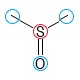}\\
\hline 
Sulfone & 
\includegraphics[width={.10\linewidth}, trim=0 0 0 -5]{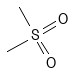} &  \includegraphics[width={.1\linewidth}]{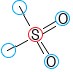} & 
Sulfonamide & 
\includegraphics[width={.11\linewidth}]{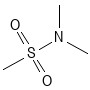} &  \includegraphics[width={.11\linewidth}]{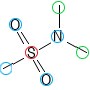}\\
\hline 
Sulfonate & 
\includegraphics[width={.11\linewidth}, trim=0 0 0 -5]{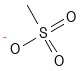} &  \includegraphics[width={.11\linewidth}]{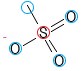} & 
Thioether & 
\includegraphics[width={.11\linewidth}]{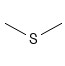} &  \includegraphics[width={.11\linewidth}]{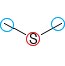}\\
\hline
Sulfate & 
\includegraphics[width={.13\linewidth}, trim=0 0 0 -5]{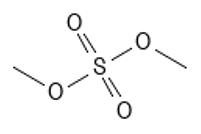} &  \includegraphics[width={.13\linewidth}]{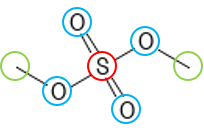} \\
\bottomrule
\end{tabular}
\end{table}

\clearpage

\newpage
\subsection{Training details}
\label{appendix:training}
In this section, we provide the data and training details. 

\textbf{Data details.} 
The tasks that are carried out includes five classification (Tox21, ClinTox, SIDER, BBBP and BACE) and three regression (ESOL, FreeSolv and Lipophilicity) tasks. The dataset descriptions are given as follows:

\begin{itemize}[leftmargin=*]
\item The Toxicology in the 21st Century (Tox21) \cite{molnet:2018, tox21:2015} dataset was created to assess the potential of drugs to disrupt biological pathways that may result in toxicity. It contains 8014 compounds with 12 different targets, which include the nuclear receptor (NR) and stress response (SR) pathways. NRs are one of the essential classes of transcriptional factors that play a critical role in human development, metabolism and physiology. The inappropriate activation of NRs can lead to a broad spectrum of negative health effects. 
SR pathways are the cellular and organismal mechanisms that resist the effects of cellular stress, which can lead to apoptosis. 

\item The ClinTox dataset compares the drugs that are approved by the FDA and drugs that have failed the clinical tests \cite{molnet:2018}. It contains 1491 drug compounds with 2 tasks: the clinical trial toxicity and the FDA approcal status. 

\item The Side Effect Resource (SIDER) \cite{sider} dataset contains information on marketed medicines and their adverse drug reactions. It contains 1427 approved drugs with 27 tasks that are grouped based on their side effects on 27 system organ classes following the MedDRA classifications \cite{meddra}.  

\item The Blood-brain barrier penetration (BBBP) dataset was created to model and predict the barrier permeability. It contains over 2000 compounds with 1 task that evaluates the permeability of the compounds. The Blood-Brain Barrier separates the circulating blood and the brain extracellular fluid, and protects the brain from foreign substances in the blood that may damage the brain. Hence, BBBP is one of the key factors in chemical toxicological studies and in drug design. 

\item The Beta-Secretase 1 (BACE) dataset consists of small molecule inhibitors, and provides quantitative IC50 and qualitative binding results for a set of inhibitors of human beta-secretase 1 (BACE1). The BACE dataset contains 1522 compounds with 1 task that provides the binding results for the inhibitors. BACE is essential for the production of the toxic amyloid beta that is critical in early part in the Alzheimer's disease pathogenesis. 

\item The Estimated Solubility (ESOL) dataset contains information on the water solubility of compounds. The ESOL dataset has 1128 compounds. The knowledge of thermodynamic solubility of drug candidates is important in drug discover as it is related to the drug absorption in the body \cite{soluble:1997}. 

\item The Free Solvation Database (FreeSolv) provides the calculated hydration free energy of fragment-like compounds in the water and provides a test of potential relevance to the binding affinity calculations for drug discovery \cite{molnet:2018}. FreeSolv contains 643 compounds. 


\item The Lipophilicity dataset \cite{molkit:2021} was used for the prediction of octanol/water distribution coefficient (logD at pH 7.4). It contains 4200 compounds. The lipophilicity of drug molecules affects both the membrane permeability and solubility.  

\end{itemize}
We use the {\fontfamily{lmtt}\selectfont AtomFeaturizer} and {\fontfamily{lmtt}\selectfont BaseBondFeaturizer} of {\fontfamily{lmtt}\selectfont DGL-LifeSci} to extract the features from the initial atom and bond features. The hypergraphs are constructed using {\fontfamily{lmtt}\selectfont DGL} and {\fontfamily{lmtt}\selectfont Networkx}.
The dataset information are given in Tables \ref{appent:bmdata}, \ref{appent:atomfeat} and \ref{appent:bondfeat}. 

\begin{table}[ht]
\caption{Datasets types, number of tasks, performance metric and split type}
\centering
\begin{tabular}{ccccc} 
\toprule
Dataset & Task & Number of tasks & Metric & Split  \\
\toprule Tox21 & Classification & 12 & AUROC & Random  \\
\hline ClinTox & Classification & 2 & AUROC & Random \\
\hline SIDER & Classification & 27 & AUROC & Random  \\
\hline BBBP & Classification & 1 & AUROC & Random  \\
\hline BACE & Classification & 1 & AUROC & Random  \\
\hline ESOL & Regression & 1 & RMSE & Random  \\
\hline FreeSolv & Regression & 1 & RMSE & Random  \\
\hline Lipophilicity & Regression & 1 & RMSE & Random  \\
\bottomrule
\end{tabular}
\label{appent:bmdata}
\end{table}

\begin{table}[ht]
\caption{Atom features used to featurize the node features}
\centering
\begin{tabular}{cc} 
\toprule
Atom Features & Number of Features \\
\toprule atom type one hot & 43 \\
\hline atomic number & 1 \\
\hline atom mass & 1 \\
\hline atom degree one hot & 11 \\
\hline atom explicit valence one hot & 6 \\
\hline atom implicit valence one hot & 7 \\
\hline atom total num H one hot & 5 \\
\hline atom formal charge one hot & 5 \\
\hline atom hybridisation one hot & 5 \\
\hline atom num radical electrons one hot & 5 \\
\hline atom is aromiatic one hot & 2 \\
\hline atom is in ring one hot & 2 \\
\hline atom chiral tag one hot & 4 \\
\hline atom chirality type one hot & 2 \\
\hline atom is chiral center & 1 \\
\bottomrule
\end{tabular}
\label{appent:atomfeat}
\end{table}

\begin{table}[ht]
\caption{Bond features used to featurize the edge features}
\centering
\begin{tabular}{cc} 
\toprule
Bond Features & Number of Features \\
\toprule bond type one hot & 4 \\
\hline bond is in ring & 1 \\
\hline bond is conjugated one hot & 2 \\
\bottomrule
\end{tabular}
\label{appent:bondfeat}
\end{table}

\newpage

The training details can be found in Table \ref{appent:molhgcn0} and \ref{appent:molhgcnext}.

\begin{table}[ht]
\centering
\caption{Hyperparameters for {\fontfamily{lmtt}\selectfont MolHMPN-NoMod}} 
\resizebox{\columnwidth}{!}{
\begin{tabular}{lcccccc}
\toprule
Dataset & $x_k$ & Cycles & GNN dropout & Regressor dropout & MLP neurons & Latent dimensions \\
\toprule Tox21 & ZERO & FALSE & $0.2$ & $0.2$ & $[64]$ & $128$ \\
\hline ClinTox & ZERO & FALSE & $0.3$ & $0.3$ & $[64, 32]$ & $128$ \\
\hline SIDER & MEAN & FALSE & $0.0$ & $0.1$ & $[64]$ & $128$ \\
\hline BBBP & MEAN & FALSE & $0.0$ & $0.0$ & $[128]$ & $256$ \\
\hline BACE & MEAN & TRUE & $0.2$ & $0.0$ & $[64,32]$ & $128$ \\
\hline ESOL & MEAN & TRUE & $0.0$ & $0.0$ & $[128]$ & $256$ \\
\hline FreeSolv & MEAN & FALSE & $0.4$ & $0.4$ & $-$ & $128$ \\
\hline Lipophilicity & MEAN & FALSE & $0.2$ & $0.2$ & $-$ & $128$ \\
\bottomrule
\end{tabular}
\label{appent:molhgcn0}
}
\end{table}

\begin{table}[ht]
\centering
\caption{Hyperparameters for MolSoft and {\fontfamily{lmtt}\selectfont MolHMPN}-0,1,2,3} 
\resizebox{\columnwidth}{!}{
\begin{tabular}{lccccccc}
\toprule
Dataset & $x_k$ & Cycles & GNN dropout & Theta dropout & Regressor dropout & MLP neurons & Latent dimensions \\
\toprule Tox21 & ZERO & FALSE & $0.2$ & $0.2$ & $0.2$ & $[64]$ & $128$ \\
\hline ClinTox & ZERO & FALSE & $0.3$ & $0.3$ & $0.3$ & $[128]$ & $256$ \\
\hline SIDER & MEAN & FALSE & $0.0$ & $0.0$ & $0.1$ & $[64]$ & $128$ \\
\hline BBBP & MEAN & FALSE & $0.0$ & $0.0$ & $0.0$ & $[128, 64]$ & $256$ \\
\hline BACE & MEAN & TRUE & $0.0$ & $0.0$ &  $0.0$ & $[128]$ & $256$ \\
\hline ESOL & MEAN & TRUE & $0.0$ & $0.0$ & $0.0$ & $[128]$ & $256$ \\
\hline FreeSolv & MEAN & FALSE & $0.4$ & $0.4$ & $0.4$ & $-$ & $128$ \\
\hline Lipophilicity & MEAN & FALSE & $0.2$ & $0.2$ & $0.2$ & $-$ & $128$ \\
\bottomrule
\end{tabular}
\label{appent:molhgcnext}
}
\end{table}

\clearpage
\newpage
\label{appendix:ablation}
In this section, we provide the details of the hyperedge construction of MolSoft.

\begin{figure}[h]
\centering
\begin{subfigure}{0.25\textwidth}
  \centering
  \includegraphics[width=0.7\linewidth]{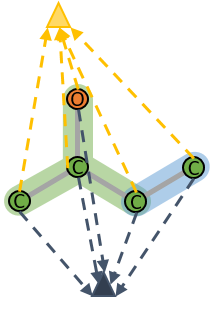}
  \caption{Before attention}
  \label{fig:sub1}
\end{subfigure}%
\begin{subfigure}{0.25\textwidth}
  \centering
  \includegraphics[width=0.7\linewidth]{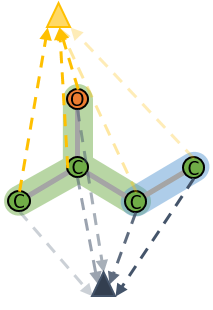}
  \caption{After attention}
  \label{fig:sub2}
\end{subfigure}

\caption{\textbf{MolSoft hypergraph construction.} The green and blue shaded parts in the molecule are the hyperedges defined in {\fontfamily{lmtt}\selectfont MolHMPN-NoMod}. We set the number of hyperedges in MolSoft to be the same as that in {\fontfamily{lmtt}\selectfont MolHMPN-NoMod}. To let the model learn the substructures, we assign all nodes in a molecule to each hyperedge as shown in a), where each dashed-arrow represents the edge that connects each node to each hyperedge. After the attention mechanism has been employed, the edges that connects the nodes to hyperedges are reweighed as shown in b).}
\label{fig:fgextract}
\end{figure}

\end{document}